\documentclass[12pt, a4paper]{article} 

\usepackage{setspace}
\usepackage[small, font=onehalfspacing]{caption}

\usepackage[authoryear]{natbib}
\bibpunct{(}{)}{;}{a}{,}{,}
\usepackage[english]{babel}
\usepackage[latin1]{inputenc}
\usepackage[T1]{fontenc}
\usepackage{lmodern}
\usepackage{amssymb}
\usepackage{xcolor,graphicx,xspace,caption}
\usepackage{listings,eso-pic,subfigure}
\usepackage{lscape}
\usepackage{multirow}
\usepackage{enumerate}
\usepackage{amsmath}
\definecolor{hint}{RGB}{191,63,0}

\definecolor{hellgelb}{rgb}{1,1,0.8}
\definecolor{colKeys}{rgb}{0,0,1}
\definecolor{colIdentifier}{rgb}{0,0,0}
\definecolor{colComments}{rgb}{1,0,0}
\definecolor{colString}{rgb}{0,0.5,0}
\def\eqdef{\stackrel{\rm def}{=}}
\usepackage{hyperref}
\hypersetup{linkcolor={blue}}

\newcommand{\E}{\mathop{\mbox{\sf E}}}

\newcommand{\Prob}{\mathop{\mbox{\sf P}}}

\newcommand{\R}{\mathbb{R}}

\newcommand{\corr}{\mathop{\hbox{Corr}}}

\newcommand{\diag}{\mathop{\hbox{diag}}}

\title{Portfolio Decisions and Brain Reactions via the CEAD method\footnote{The authors greatfully acknowledge financial support from the Deutsche Forschungsgemeinschaft through SFB 649 ''Economic Risk'' and IRTG 1792 ''High Dimensional Non Stationary Time Series''.}
}
\author{Piotr Majer\footnote{Humboldt-Universität zu Berlin,
C.A.S.E. - Center for Applied Statistics and Economics, Spandauer Str. 1, 10178 Berlin, Germany, tel: +49 (0)30 2093 5623, fax: +49 (0)30 2093 5649}, Peter N.C.~Mohr\footnote{Department of Education and Psychology, Freie Universit\"at Berlin, Berlin}, Hauke R.~Heekeren\footnote{Department of Education and Psychology, Freie Universit\"at Berlin, Berlin},\\ Wolfgang K.~Härdle\footnote{Humboldt-Universität zu Berlin, C.A.S.E. - Center for Applied Statistics and Economics, Spandauer Str. 1, 10178 Berlin, Germany and School of Business, Singapore Management University, 50 Stamford Road, Singapore 178899}
}
\date{\;}

\usepackage[top = 1in, bottom = 1in, left = 1in, right = 1in]{geometry}

\begin{document}
\maketitle
\begin{abstract}
\footnotesize{\noindent
Decision making can be a complex process requiring the integration of several attributes of choice options. Understanding the neural processes underlying (uncertain) investment decisions is an important topic in neuroeconomics. We analyzed functional magnetic resonance imaging (fMRI) data from an investment decision study for stimulus-related effects. We propose a new technique for identifying activated brain regions: Cluster, Estimation, Activation and Decision (CEAD) method. Our analysis is focused on clusters of voxels rather than voxel units. Thus, we achieve a higher signal to noise ratio within the unit tested and a smaller number of hypothesis tests compared with the often used General Linear Model (GLM). We propose to first conduct the brain parcellation by applying spatially constrained spectral clustering. The information within each cluster can then be extracted by the flexible Dynamic Semiparametric Factor Model (DSFM) dimension reduction technique and finally be tested for differences in activation between conditions. This sequence of Cluster, Estimation, Activation and Decision admits a model-free analysis of the local fMRI signal. Applying a GLM on the DSFM-based time series resulted in a significant correlation between the risk of choice options and changes in fMRI signal in the anterior insula and dorsomedial prefrontal cortex. Additionally, individual differences in decision-related reactions within the DSFM time series predicted individual differences in risk attitudes as modeled with the framework of the mean-variance model. 
}
\bigskip

\noindent This is a post-peer-review, pre-copyedit version of an article published in Psychometrika Vol. 81, No. 3, 881-903 (2016). The final authenticated version is available online at \href{http://dx.doi.org/10.1007/s11336-015-9441-5}{doi: 10.1007/s11336-015-9441-5}.
\end{abstract}

\pagebreak

\section{Introduction}\label{intro}

Economic decision making takes place when, for example, an individual buys beverages in a supermarket, purchases a car or chooses an investment fund. Some of these choices are made when the outcome is uncertain and hard to anticipate, which is particularly true for an investment decision. The decision-making process builds on different mechanisms such as representation and integration of relevant evidence for and a comparison process of different choice options. This mechanism has attracted considerable attention in many different fields, from cognitive psychology, behavioral economics, to neuroscience, see, e.g., \cite{neurobook}. Economic decisions are usually explained in a value-based scheme, where different choice options are evaluated and the option with the highest value is chosen. The values attributed to different incarnations of options may be generated by a nonobservable utility function. It was first formalised by \cite{bern} and further developed by \cite{neumann:1953} and \cite{kahneman:tversky:1979} to address the uncertainty of outcomes. In this case individual risk preferences are attributed to the curvature of the utility function. Alternatively, decision making can be explained in a framework of risk-return models, which incorporate the risk attitude as a weighting factor, see, e.g., \cite{webermili}. 

Research in the field of Decision Neuroscience (as well as its sub-field Neuroeconomics) attempts to address human economic behavior (i.e., decisions) by looking at neural systems that underlie decision making (e.g., \cite{Camerer, heekeren:2008}). In practice one measures changes in brain activity using methods such as electroencephalography (EEG) and functional magnetic resonance imaging (fMRI), see, e.g., \cite{neurobook2}. FMRI is based on measuring the blood-oxygen-level-dependent (BOLD) signal and captures parameters related to changes in blood flow and blood oxygenation. FMRI data are recorded over time, for example during multiple investment decisions. The captured changes in fMRI BOLD signal are indirectly related to neural firing rates \citep{Logo2008}. The acquired images are high-dimensional and detecting stimulus-related effects is a non-trivial task. Changes in brain activation in response to decision making may be of a modest size (i.e., in comparison to reactions to visual or auditorial stimuli) and possible hemodynamic responses may be subtle and hardly detectable in the BOLD signal. It poses a genuine challenge to all existing methods and may require some extraordinary techniques.

A benchmark method to detect brain regions activated by the stimulus is the general linear model (GLM). GLM is a single-voxel technique which tests each voxel separately and results in a $3$-D map of changes in fMRI signal. The test is done in a linear regression setup, where the voxel time series are modeled according to the hypothesized and predefined regressors (design matrix), which correspond to the experimental paradigm and potential confounds. This simple methodology has proved to be extremely successful in practice and has led to a wealth of important findings (e.g., \citet{kable:glimcher:2007}), also regarding investment decisions \citep{mohr:2010, mohr:2010b}. Nevertheless, it has several limitations. Firstly, all neural activity not predefined in the design is neglected and cannot be identified by the model. In contrast to this model-based approach, recently introduced model-free approaches \citep{beckmann:smith:2005, myfmri} offer to identify effects without any a priori hypothesis. Secondly, possible information reflected in variability and higher moments of the BOLD signal \citep{mohr:nagel:2010, Garrett} is disregarded by the GLM approach. Moreover, activation maps derived by the single-voxel approach may by "inherently limited" by a typically low signal to noise ratio of individual voxel data, as reported by \cite{Heller}. Alternatively, a simultaneous analysis of multi-voxel data that co-vary with the experimental design may increase the signal without adding noise. 

To overcome these shortcomings we follow the idea of \cite{Heller} and focus our analysis on the cluster rather than voxel unit. This leads in fact to an alternative technique for analyzing fMRI data, where the brain parcellation serves as a starting point. The fMRI clustering is done by the normalized cut spectral algorithm \citep{Ncut:1997} which became very popular in neuroscience, see, e.g., \cite{Craddock:2012}. The algorithm makes use of a correlation between neighboring voxels which defines their proximity. Thus, a possible co-movement (i.e., simultaneous hemodynamic response) plays a key role in defining a homogeneous cluster. The shape and spatial structure is data-driven and clusters are contiguous volumes of voxels, ensuring interpretability. After functional connectivity maps are constructed one needs to investigate neural activity displayed by the cluster unit. Our approach is model-free, the signal carried within a cluster is extracted by the dynamic semiparametric factor model (DSFM). The DSFM, proposed by \cite{DSFM:2009}, is employed here as a dimension reduction technique \citep{myfmri}. It filters the noise and extracts only the common temporal information (i.e., joint reaction by neighboring voxels to the stimulus). The resulting simple, denoised temporal representation of cluster dynamics may be tested for activation within the GLM framework or using a model-free approach. Our technique: Cluster, Estimation, Activation and Decision (CEAD) method combines parcellation based on functional connectivity and DSFM. Thus, it greatly simplifies the complexity of the data while preserving the high accuracy of the representation. Particularly this high spatiotemporal accuracy is of great importance, when stimulus related effects may be subtle and local (such as in investment decisions under risk).

The presented methodology is applied to investigate a possible relationship between individual differences in risk preferences and dynamics in the BOLD response. In the first step the extracted temporal information from clusters is tested for changes in brain activation. These, possibly few, activated clusters correlated with risk are further investigated with respect to risk attitudes estimated from subject responses to investment decision (ID) tasks. Here, we establish a link between changes in BOLD signal and individuals' risk weights in a risk-return model. Based on this analysis we identify bilateral anterior insula (aINS) activity as a correlate of risk (standard deviation). The risk attitudes, derived from the subject's investment decisions are successfully predicted based only on underlying brain activity in aINS.

In the upcoming section (\ref{methods}) we describe the experimental procedures, our methodology and derivation of risk attitudes. At the end of that part a short simulation study of testing performance is shown. In the next section (\ref{results}) our modeling parameters and empirical findings are reported. We show and exploit the relation between risk preferences and temporal information extracted from clusters. Our conclusions are detailed in the discussion section.

\section{Materials and Methods}\label{methods}

In this section our experimental and fMRI data acquisition setup is presented. In the next step we describe our methodology and employed statistical tools. It begins with an introduction to the normalized cut spectral clustering. Secondly, the advanced dimension reduction technique: DSFM is discussed. It shows how to extract a temporal information (i.e., hemodynamic response) from entire clusters. We briefly sketch our activation testing procedure which is similar to the voxelwise GLM approach. The testing performance is evaluated in a simulation study. Finally, we introduce the risk-return model and estimate the subjects' risk attitudes based on their investment decisions.
\subsection{Experimental Procedures}
Subjects, $I=19$, performed an adjusted version of the Risk Perception in Investment Decisions Task \citep{mohr:2010}. In this task subjects see past returns of either one single investment or two investments that form a portfolio ($50\%$ of the money invested in each). While they see the past returns they have to make a choice between, if they would prefer to invest in a bond with $5\%$ fixed return or the investment that is displayed (either single risky investment or risky portfolio). The choice situations differed in three within-subject conditions: (A) choices between $5\%$ fixed return and a single risky investment, (B) choices between $5\%$ fixed return and a risky portfolio of $2$ single investments with perfectly ($\rho=1$) correlated returns, and (C) choices between $5\%$ fixed return and a risky portfolio of $2$ single investments with uncorrelated returns ($\rho=0$). Importantly, the return history of the risky options (either single investment or portfolio) was exactly the same in all $3$ conditions. All displayed returns were gaussian with different set of parameters $\mu$ and $\sigma$, where $\mu=5\%, 7\%, 9\%, 11\%$ and $\sigma=2\%, 4\%, 6\%, 8\%$. Each of the choices regarding single investments was repeated once to hold the number of choices between the bond and a single investment and the bond and a portfolio constant. In total subjects made $256$ choices in two blocks of $128$ choices each. Subjects had a maximum of $7$ seconds to enter their choices via a response box with two buttons. The location of the choice options on the screen was counterbalanced between left and right to avoid order effects.
\subsection{fMRI Data}
MRI data were acquired on a $3$ T scanner (Trio; Siemens) using a $12$-channel head coil. Functional images were acquired with a gradient echo T$2$*-weighted echo-planar sequence (TR = $2000$ ms, TE = $30$ ms, flip angle = $70$, $64 × 64$ matrix, field of view = $192$ mm, voxel size = $3\times 3\times 3$ mm$^3$). A total of $37$ axial slices ($3$ mm thick, no gap) were sampled for whole-brain coverage. Imaging data were acquired in two functional runs with $695$ and $705$ volumes respectively. A high-resolution T$2$-weighted anatomical scan of the whole brain was acquired ($256 \times 256$ matrix, voxel size = $2\times 2\times 2$ mm$^3$).

The data was initially pre-processed with FSL 4.0 (FMRIB's Software
Library). Pre-processing included motion correction and slice-time correction. Additionally, images were normalized into a standard stereotaxic space (Montreal Neurological Institute (MNI), Montreal, Quebec, Canada). As a result high-dimensional data was obtained $91\times109\times91\times1400$, where $t=1,\ldots,1400$ for each subject $i=1,\ldots,19$. 
\subsection{fMRI Analysis}
The key idea of this study is to use data-driven, contiguous clusters as the units of the analysis. The clustering is done by a Spatially Constrained Spectral Clustering algorithm which became extremely successful in neuroscience, see, e.g., \cite{Craddock:2012}. In the second step, temporal information contained in each cluster is extracted by the DSFM approach, as an alternative to averaging over voxels in the clusters proposed by \cite{Heller}. Comparison with the latter approach is presented in a simulation study (see section \ref{simu}) and our empirical results. After the cluster temporal information is extracted, activated regions of interest (ROIs) are found by the GLM testing procedure.
\subsubsection{Spatially Constrained Spectral Clustering} \label{NCUT}
The brain parcellation results from normalized cut spectral clustering (NCUT). This technique, first proposed by \cite{Ncut:1997}, is reported to be robust to outliers \citep{Luxburg:2007} and computationally efficient. It also allows for a simple incorporation of constraints, i.e., a spatial contiguousness, which can be exploited in the human brain mapping. The method was introduced to the field of cognitive neuroscience by \cite{Ncut:2008, Shen:10, Craddock:2012}. \cite{Shen:10} reported that task-related fMRI data may be analyzed with this algorithm and that the resulting brain parcellation is highly consistent with the resting-state fMRI. The NCUT approach is closely related to the graph theoretic formulation of clustering. The set of voxels $Y=(Y_1,\ldots,Y_J)$ is represented as a weighted undirected graph, where the nodes of the graph are the voxels and an edge is given between every pair of voxels $Y_j$ and $Y_{j'}$. The weight on each edge, denoted by $w(j,j')$, is a proximity measure between voxels (nodes) $j$ and $j'$, and is defined as in the previous paper: 
\begin{align}
w(j,j') = \left\{\begin{array}{ll}
\text{max}\left\{\corr_{t}(Y_{j},Y_{j'}),0\right\} & \text{, for} \left\|X_j-X_{j'}\right\|<d,\\
0 & \text{, otherwise,}
\end{array}\right.
\label{eq:similarity}
\end{align}
where $\left\|\cdot\right\|$ denotes the Euclidean norm in $\R^3$ space, $X_j \in \R^3$ are $j$-th voxel coordinates. The radius $d$ is selected in such a way that only the $26$ nearest neighbors (face and edge touching; $3$-D neighborhood of a single voxel) are included. Such a constraint ensures a contiguous shape of each cluster \citep{Xu:2005, Kamvar}. Moreover, the similarity matrix $W=\left\{w(j,j')\right\}_{j,j'=1,\ldots,J}$ (of size $J\times J$) derived by (\ref{eq:similarity}) is sparse and thus computational complexity is reduced.
The similarity between voxels in $3$-D neighborhood is given by correlation coefficient of the voxels time series with a threshold to make it non-negative. By applying the correlation as a similarity measure we ensure the temporal homogeneity within a cluster, which is further exploited in the next section (\ref{dsfm}). Once a proximity measure is chosen, a group-building algorithm for creating a functional connectivity map needs to be specified. The NCUT algorithm is a hierarchical procedure, it starts with the coarsest partition possible: one cluster contains all of the voxels. It proceeds by splitting the single cluster up into smaller sized clusters until a pre-specified number of groups $C$ is achieved. The partition of an initial set is done such that the similarity between voxels within the proposed group is greater than the similarity between voxels in different groups. For example, for two disjoint groups $P$ and $Q$, one computes the normalized cut cost by:
\begin{equation}
N_{cut}(P,Q)=\frac{\sum_{Y_j\in P, Y_{j'}\in Q}w(j,j')}{\sum_{Y_j\in P, Y_{j'}\in R}w(j,j')}+\frac{\sum_{Y_{j}\in P, Y_{j'}\in Q}w(j,j')}{\sum_{Y_j\in Q, Y_{j'}\in R}w(j,j')},
\label{eq:Ncut}
\end{equation}
where $R=Q+P$ is the initial set that has to be partitioned. The denominators in the formula (\ref{eq:Ncut}) may be seen as a sum of all similarities between sets $P$ and $Q$ that are neglected in this division. The nominators stand for all the similarities between the proposed groups ($P$ and $Q$) and the initial set $R$, thus a size of a group has an influence on the normalized cut cost. Finding an optimal division of set $R$ might be found by minimizing the normalized cut criterion:
 \begin{equation}
(P^*,Q^*)=\arg\min_{R=P+Q} N_{cut}(P,Q).
\end{equation}
 Therefore we ensure that, simultaneously, similarities within each cluster are maximized and similarities between clusters are minimized.
 This approach leads to balanced sizes of clusters and reduces the likelihood of obtaining singletons as a result.
\cite{Ncut:1997} showed that minimizing (\ref{eq:Ncut}) is equivalent to minimizing the Rayleigh quotient denoted by:
\begin{equation}
Q(y)=\frac{y^{\top}\mathcal{L}y}{y^{\top}Dy},
\label{eq:NcutQ}
\end{equation}
under the constraint that $y$ is a piecewise (discrete) vector $J\times 1$ and $y^{\top}\diag(D){1}_{J}=0$. Matrix $\diag(D)$ is defined by $D=(d_1,\ldots,d_J)$ a degree vector, $d_j=\sum^{J}_{j'=1}w(j,j')$ and $\mathcal{L}$ is the Laplacian of the graph given by:
\begin{align}
\mathcal{L}(j,j')=\left\{\begin{array}{lll} d_{j} & ,\ j=j',\\
-w(j,j') & ,\  w(j,j')>0,\\
0 & ,\ \text{elsewhere.}   \end{array}\right.
\label{eq:laplacian}
\end{align}     
Minimizing the formula (\ref{eq:NcutQ}) is closely related to spectral clustering, where the first nontrivial eigenvector of the graph Lapacian matrix $\mathcal{L}$ is used. The authors showed that the problem is NP-complete, an approximate discrete solution can be found efficiently.
\subsubsection{Dynamic Semiparametric Factor Model}\label{dsfm}
The clusters are constructed to maximize the temporal homogeneity between voxels. Their similar time evolution (i.e., reflected in joint hemodynamic response after stimuli) explicitly suggest possible low-dimensional representation of the multidimensional time series. The temporal variability in the cluster
series, that may be related to investment decisions and possibly individual differences in risk
attitude, is captured by a dynamic semiparametric factor model (DSFM), proposed by \cite{DSFM:2009}. DSFM serves here as a dimension reduction technique, which is able to extract temporal dynamics from the functional connectivity brain maps by corresponding low dimensional time series (factor loadings) in
only one estimation step. Due to a subject-specific spatial structure of the brain functional connectivity maps, we model each cluster separately.

The BOLD signal of all voxels in a single cluster $c$, $c=1,\ldots,C$
during the entire experiment is a multi-dimensional
time series. The stated below DSFM is designed to model
such high-dimensional time series:
\begin{eqnarray} \label{eq:DSFM}
 Y_{t,j}&=&m_0(X_{t,j}) + \sum_{l=1}^L Z_{t,l}\ m_l(X_{t,j})+
 \varepsilon_{t,j}, \;\; 1 \le j \le J_c,  \; 1 \le t \le T. \nonumber\\
 &\eqdef& Z^\top_t m(X_{t,j}) +  \varepsilon_{t,j} = Z_{t}^{\top} A^* \Psi_{t,j} +  \varepsilon_{t,j},
\end{eqnarray}
where $Z_t = (\textbf{1}, Z_{t,1}, \ldots, Z_{t,L})^{\top}$ is a
latent $(L+1)$-dimensional stochastic process and $m$ is an
$(L+1)$-tuple $(m_0,\ldots,m_L)$ of unknown real-valued functions
$m_l$. More precisely, the voxel's coordinates $(x_1, x_2, x_3)\in \R^3$ that belongs to an analyzed cluster $c$ is the
covariate $X_{t,j}$ (in this setup it is time-invariant $X_{t,j}=X_{j}$) and the normalized BOLD signal is the dependent
variable $Y_{t,j};\;j=1,\dots,J_c;\; t=1,\dots,T$. We assume  $\varepsilon_{t,j} \bot Z_{t,j}$, $\E\varepsilon_{t,j}=0$ and $\E\varepsilon_{t,j}^2<\infty$. The functions $m_l$ are given as a linear combination of space basis functions
$\Psi_{t,j} =[\psi_{1}(X_{t,j}),\dots,\psi_{K}(X_{t,j})]^\top$ and
corresponding $(L+1) \times K$ matrix of unknown coefficients
$A^{\ast}$. In our setup,
$[\psi_{1}(X_{t,j}),\dots,\psi_{K}(X_{t,j})]^\top$ are quadratic
tensor B-splines on $K$ equidistant knots. To find the estimates of
$Z_{t}^{\top}$ and $A^{\ast}$ one solves:

\begin{equation}
(\widehat{Z}_t, \widehat{A^*}) = \arg\min_{Z_t,A^{\ast}}
\sum^{T}_{t=1}\sum^{J}_{j=1}\{Y_{t,j} - Z_t
A^{\ast}\Psi_{t,j}\}^2\;.
\label{eq:NR}
\end{equation}
A solution to the problem stated in (\ref{eq:NR}) may be found by the Newton-Raphson method. Time dynamics are represented by $\widehat{Z}_t$, while
$\widehat{A^*}$ captures the smooth, nonparametrically estimated spatial structure of clusters.

In the formula (\ref{eq:DSFM}) the time frame is constant over all clusters and equals $T=1400$. Due to varying spatial structure and size of each cluster $c$, $c=1,\ldots,C$, we denote the dimension $J_c$ as the $c$ cluster size. The statistical inference of the each cluster is then based on the
low-dimensional time series analysis for ${Z}_{t}$. As shown by \cite{DSFM:2009}, the inference based on the estimates $\widehat{Z}_{t}^{\top}$ holds for ``true'' unobserved time series $Z_{t}^{\top}$, as the difference between $Z_{t}^{\top}$  and $\widehat{Z}_{t}^{\top}$ is asymptotically negligible.
\subsubsection{General Linear Model and Testing Procedure}\label{GLM}
In practice, the analysis of BOLD fMRI data is conducted using voxelwise General Linear Model, see, e.g, \cite{friston95} and \cite{Worsley2002}, where the magnetic resonance signal at voxel $j$ is modeled by:
\begin{equation}
Y_j=\widetilde{X}\beta_j+e_j,
\end{equation}
where $\widetilde{X}$ denotes the $T\times p$ design matrix, $\beta_j$ is the $p \times 1$ vector of regression coefficients and $e_j$ is a (often serially correlated) measurement error. The matrix $\widetilde{X}$ is constructed as a convolution of hemodynamic response function (HRF) $h(t)$ and the stimulus time signal and might also incorporate additional elements (i.e., temporal derivatives) when required by a specific experiment setup. It is common practice to model the HRF by a difference of two gamma functions, i.e., 
\begin{equation*}\displaystyle h(t)=(\frac{t}{5.4})^{6}\exp\left\{-(t-5.4)/0.9)\right\}-0.35(\frac{t}{10.8})^{12}\exp\left\{-(t-10.8)/0.9\right\},
\end{equation*} 
see, e.g., \cite{Worsley2002}. Inference focuses on the estimates $\widehat{\beta}_j$ and the hypothesis $H_0 : \beta_j=0$ is tested voxelwise (first-level analysis). $\widehat{\beta}_j$ being significantly different from $0$ is interpreted as activation at the voxel $j$. Group analysis is usually done in the mixed-effects framework, where the activation pattern for $i$ subject at $j$ voxel $\widehat{\beta}_j^i$ serves as an input for the model (higher-level analysis). This standard technique implemented in FSL's FLAME (FMRIB's local analysis of mixed effects) is used here to test whether regression coefficients are significant and activation can be reported at the group level. The region of interest is reported to be significantly activated for clusters reaching uncorrected threshold of $Z$-score $>3.09$ and consisting of at least $20$ neighboring voxels. For more details we refer here to the technical reports of the FMRIB Analysis Group, see, e.g., \cite{featglm1} and \cite{featglm2}.
\subsubsection{Cluster, Estimation, Activation and Decision (CEAD) Method}
The resulting cluster representation by $\widehat{Z}_{t}^{\top}$ serves as the unit of analysis for the relevant signals related to the ID tasks and decisions. Profiting from higher signal-to-noise ratio present on the group level \citep{Heller} clusters are tested for activation. For analysis of all participated subjects $i=1,\dots,I$, our multivariate scheme may be
summarized in the following steps:
\begin{enumerate}
\item \textbf{C}luster-step: for each subject $i$ construct the brain parcellation into $C$ groups using spectral clustering NCUT algorithm.
\item \textbf{E}stimation-step: given the subject-specific clustering results, for subject $i$ take the $c$ cluster and fit the DSFM, given in (\ref{eq:DSFM}). Repeat this estimation procedure for all clusters $c=1\dots,C$ and all subjects $i = 1,\ldots,I$. The DSFM approach is thus applied $C \times I$ times separately.
\item \textbf{A}ctivation-step: representing $(i,c)$, $i = 1,\ldots,I$, $c=1\dots,C$ cluster dynamics by low-dimensional representation $\widehat{Z}^{(i,c)}_{t}$ test the time series activation in the GLM framework. Select the activated clusters that are related to neural processes underlying (risky) investment decisions.
\item \textbf{D}ecision-step: investigate the activated factor loadings $\widehat{Z}^{(i,c)}_{t}$. Is the subjects investment behavior represented in any of the activated clusters? Is there any relation between the risk attitude and the low-dimensional time series?
\end{enumerate}
\subsection{Simulation Study}\label{simu}
\begin{figure}[h]
\begin{center}
    \includegraphics[height=2.3cm]{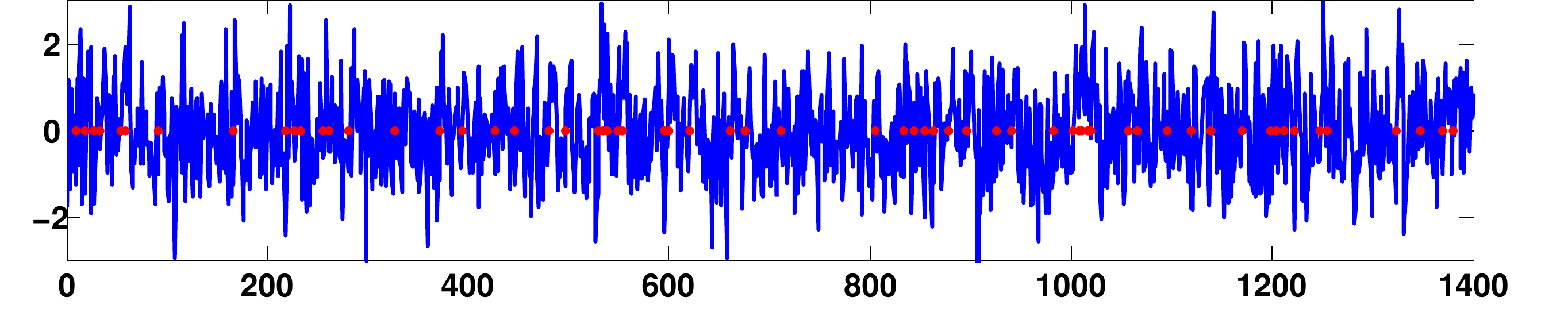}\\
		    \includegraphics[height=2.3cm]{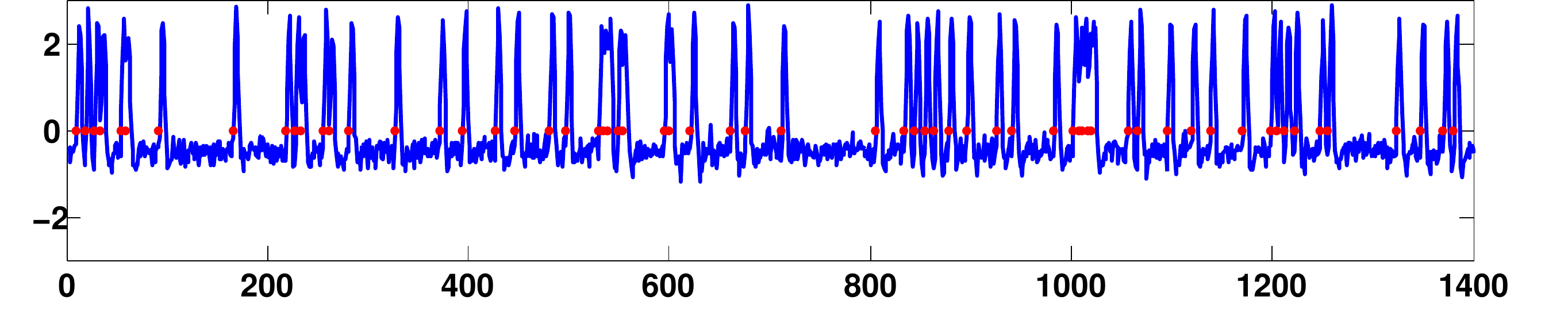}
				\end{center}
				\vspace{-0.3cm}
        \caption{Setup (a): the simulated  $(1,1,1)$ voxel $Y_{t,1}$ (top) and the estimated $\widehat{Z}_t$ (bottom) plotted against time (each $2$ seconds); red dots denote stimulus; $\corr_t(\widehat{Z}_t,\text{stimulus})=0.98$.}
				\label{fig:simua}
\end{figure}
\begin{figure}[!h]
\begin{center}
    \includegraphics[width=0.95\textwidth]{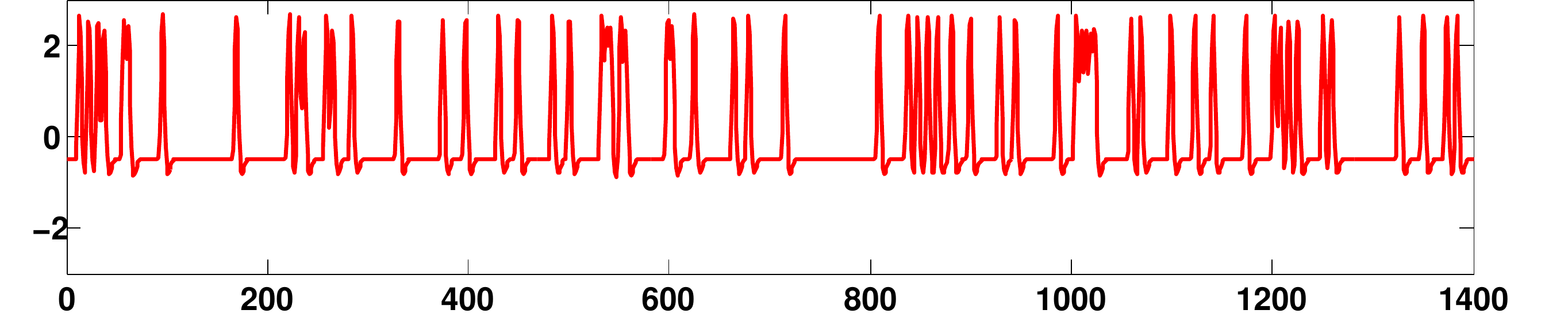}
		\end{center}
        \caption{Stimulus time series derived as a convolution of double Gamma hemodynamic response function and uncorrelated portfolio stimulus $\times 64$ plotted against time (each $2$ seconds).}
				\label{fig:stimul}
\end{figure}
\begin{figure}[!h]
\begin{center}
    \includegraphics[width=0.95\textwidth]{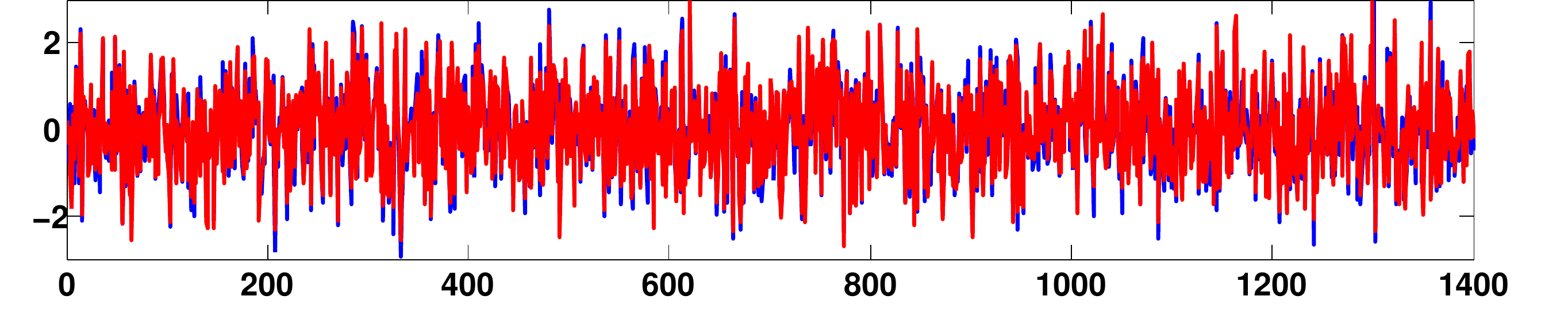}
		\vspace{-0.2cm}
        \caption{Simulated spatially correlated Gaussian noise for $2$ vertical neighbor voxels (red and blue) plotted against time (each $2$ seconds); $\corr_t(\varepsilon_{t,1},\varepsilon_{t,2})=0.97$ .}
				\label{fig:corrnoise}
				\end{center}
\end{figure}

This part of our study is designed to investigate the performance of the proposed method in a simulation study. Our approach is evaluated against the benchmark, voxelwise GLM and the averaging technique introduced by \cite{Heller} (in each cluster take average over voxels and test for activation). We simulated data at one, exemplary cluster on the $6\times 7 \times 6$ grid that mimics the average cluster obtained in our empirical analysis: $Y_t=Z_{t}^{\top}m({X})+\varepsilon_{t}$, where $Y_t$ is a $6\times 7 \times 6 \times 1400$ BOLD signal, $m(X)=m(x,y,z)=\left\|(x,y,z)-(6,8,6)\right\|$ is a smooth spatial structure, $Z_{t}$ is a (perfect) stimulus time series (HRF $\times 64$, see Figure \ref{fig:stimul}) and $\varepsilon_t$ is noise. The (single) factor $m(\cdot)$ is a smooth, non-linear function that decreases in the direction of the point (6,8,6), that is not present on the grid, thus $m(\cdot)>0$. The $Z_{t}$ is the simplest design matrix (here $1\times 1400$) from GLM setup and in this case stands for all stimuli corresponding to the correlated portfolio from our experiment. Therefore, we assume that only one true neural process is present in this cluster. We investigate two possible cases for $\varepsilon_t$ ($6\times 7 \times 6 \times 1400$): (a) $\varepsilon_t$ is i.i.d. Gaussian and (b) $\varepsilon_t$ is spatially correlated Gaussian; $\mu=0$ and $\sigma=1$. The spatially correlated noise time series $\varepsilon_{sc,t}$ is derived (independently at each $t$, $t=1,\ldots,1400$) as a convolution of i.i.d. Gaussian noise from (a) with a spatial Gaussian kernel (FWHM $8$ mm) and depicted in Figure \ref{fig:corrnoise}.
\begin{figure}
\begin{center}
    \includegraphics[height=2.3cm]{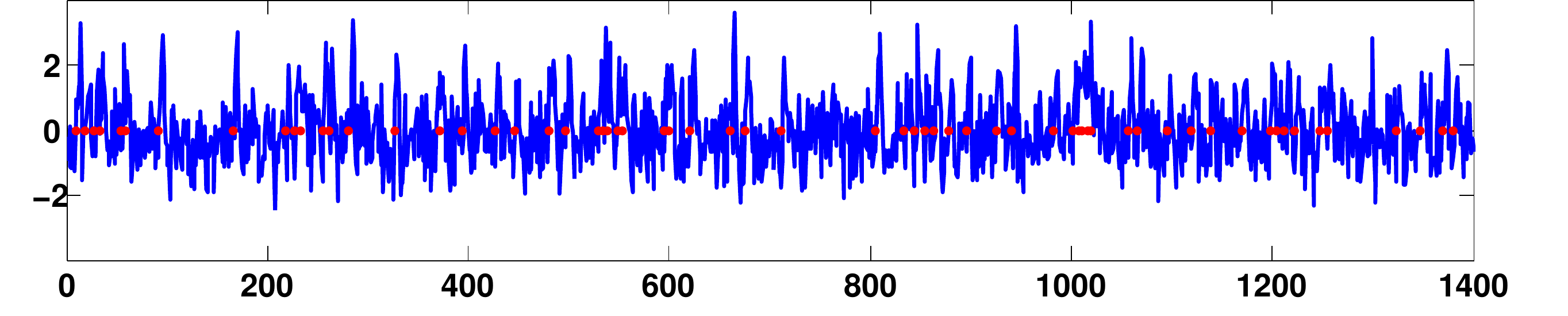}\\
		    \includegraphics[height=2.3cm]{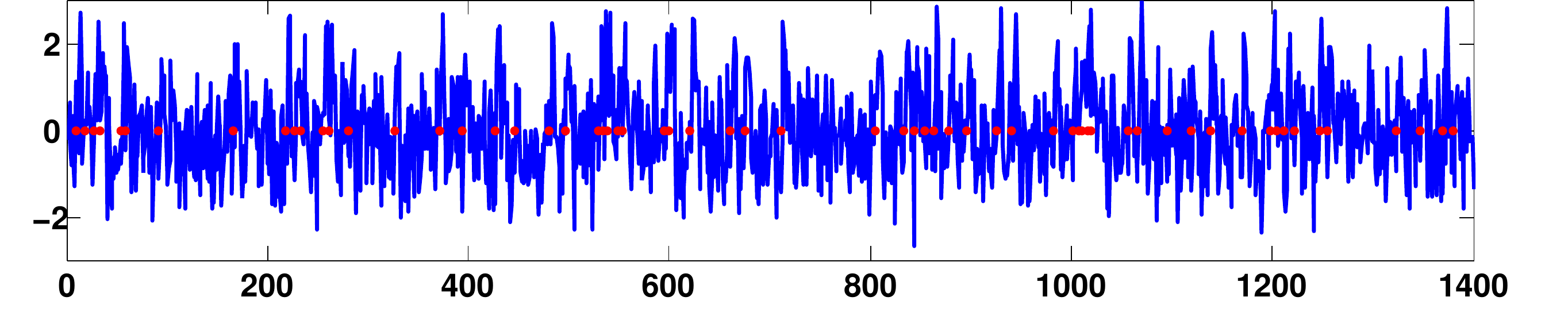}
		\end{center}
		\vspace{-0.3cm}
        \caption{Setup (b): the simulated $(1,1,1)$ voxel $Y_{t,1}$ (top) and the estimated $\widehat{Z}_t$ (bottom) plotted against time (each $2$ seconds); red dots denote stimulus; $\corr_t(\widehat{Z}_t,\text{stimulus})=0.60$.}
				\label{fig:simub}
\end{figure}
Examples of simulated BOLD signals are shown in Figures \ref{fig:simua} and \ref{fig:simub}. The performance for all three techniques: DSFM with $L=1$, GLM (pre-smoothed with FWHM $8$ mm) and averaging over voxels in the cluster (with and without pre-smoothing) for the setup (a) is remarkably good and all statistics are higher than $100$. The (b) study is summarized in Table \ref{tab:simu}. Firstly, all investigated techniques discover a significant activation and yield similar results. Secondly, the maximum $Z$-score in the GLM approach is the highest test statistics in all cases. When the $Z$-scores are averaged over all voxels, the DSFM approach yields the best result. Moreover, the simple averaging approach is outperformed by the DSFM. We conclude, that DSFM might serve as an interesting alternative to the benchmark GLM method, especially if the analysis goes beyond an identification of activation patterns (i.e., higher moments, time series analysis of voxels in a neighborhood).
\begin{table}[h]
\begin{center}
\begin{tabular}{r r@{.}l r@{.}l r@{.}l r@{.}l r@{.}l r@{.}l r@{.}l}
\hline\hline
 & \multicolumn{2}{c}{GLM} & \multicolumn{2}{c}{DSFM} & \multicolumn{2}{c}{Average(s)} & \multicolumn{2}{c}{Average} \\
 \hline
max $Z$-score & $30$ & $54$ & $27$ & $96$ & $27$ & $14$ & $27$ & $48$\\
mean $Z$-score & $26$ & $34$ & $27$ & $96$ & $27$ & $14$ & $27$ & $48$\\
\hline\hline
\end{tabular}
\end{center}
\caption{Test statistics $Z$-scores derived in simulation setup (b) for GLM, DSFM, averaging and averaging for smoothed (FHWM 8mm) data denoted by Average(s).}
\label{tab:simu}
\end{table}

\begin{figure}[!h]
\begin{center}
    \includegraphics[width=0.95\textwidth]{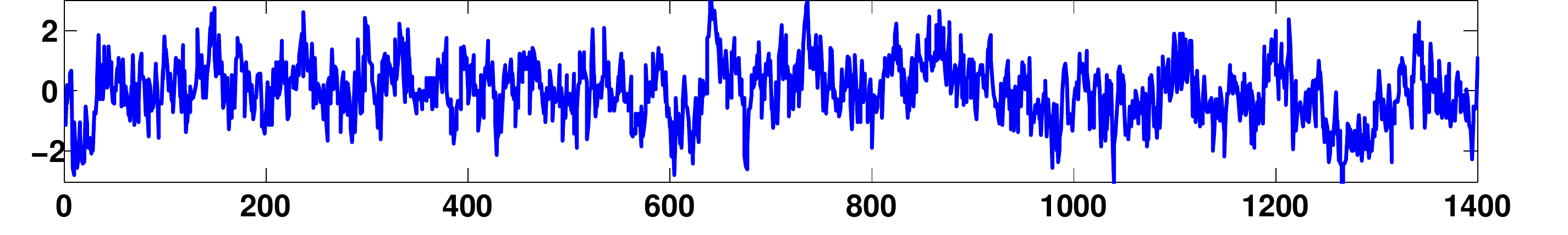}
		\end{center}
        \caption{Simulated stimulus time series as the AR($2$) process: $\widetilde{Z}_{t}=0.5\widetilde{Z}_{t-1}+0.2\widetilde{Z}_{t-2}+\varepsilon_{AR,t}$, plotted against time (each $2$ seconds).}
				\label{fig:stimul2}
\end{figure}

The performance of the proposed method is also studied, when the exemplary cluster does not exhibit stimulus-related effects. In particular, we simulated the $6\times 7 \times 6 \times 1400$ BOLD signal $Y_t=\widetilde{Z}_{t}^{\top}m({X})+\varepsilon_{sc,t}$, where: (c) $\widetilde{Z}_{t}={1}_{1400}$ is a constant series of ones and (d) $\widetilde{Z}_{t}$ is a simulated autoregressive process of order $2$, where $\widetilde{Z}_{t}=0.5\widetilde{Z}_{t-1}+0.2\widetilde{Z}_{t-2}+\varepsilon_{AR,t}$, $\varepsilon_{AR,t}$ is a white noise independent of $\varepsilon_{sc,t}$ and $\corr_t(\widetilde{Z}_t,\text{stimulus})=0.04$, see Figure \ref{fig:stimul2}. Therefore, the setup (c) corresponds to a case, when only the (spatially correlated) noise is present in the cluster and there is no common neural signal. The setup (d) assumes a common neural process which is not related to the stimulus. The results of all $3$ techniques are summarized in Table \ref{tab:simu1}. The resulting $Z$-scores are remarkably smaller than a typical threshold $3.09$ and the stimulus-related effects are not identified. Furthermore, all approaches yield similar results.   
\begin{table}[h]
\begin{center}
\begin{tabular}{r r@{.}l r@{.}l r@{.}l r@{.}l r@{.}l r@{.}l r@{.}l}
\hline\hline
 & \multicolumn{2}{c}{GLM} & \multicolumn{2}{c}{DSFM} & \multicolumn{2}{c}{Average(s)} & \multicolumn{2}{c}{Average} \\
 \hline
max $Z$-score & $1$ & $90$ & $0$ & $38$ & $0$ & $68$ & $0$ & $62$\\
mean $Z$-score & $0$ & $61$ & $0$ & $38$ & $0$ & $68$ & $0$ & $62$\\
\hline
max $Z$-score & $1$ & $66$ & $1$ & $10$ & $1$ & $07$ & $1$ & $10$\\
mean $Z$-score & $0$ & $99$ & $1$ & $10$ & $1$ & $07$ & $1$ & $10$\\
\hline\hline
\end{tabular}
\end{center}
\caption{Test statistics $Z$-scores derived in simulation setup (c)-upper and (d)-lower panel, respectively, for GLM, DSFM, averaging and averaging for smoothed (FHWM 8mm) data denoted by Average(s).}
\label{tab:simu1}
\end{table}

\subsection{Behavioral Modeling}

The subject specific risk attitudes can be directly derived from subject responses to ID tasks. Following \cite{markowitz:1952, caraco80} we apply the benchmark - mean-variance model to reflect the subjects decision making process:
\begin{equation}
V_i(x) =  \overline{x} - \phi_i S(x),
\end{equation}
where $V_i(x)$ is the \emph{value} a subject $i$
assigns to an investment $x$, $\overline{x}$ is an empirical mean and represents the
\emph{expected return},  $S(x)$ stands for a standard deviation and represents the
subject's \emph{risk}, and $\phi_i$ is the individual risk
weight: risk attitude. Therefore, in line with the portfolio theory introduced by \cite{markowitz:1952}, we follow the common mean-variance approach. 

\begin{figure}[ht]
    \begin{center}
    \includegraphics[width=10cm]{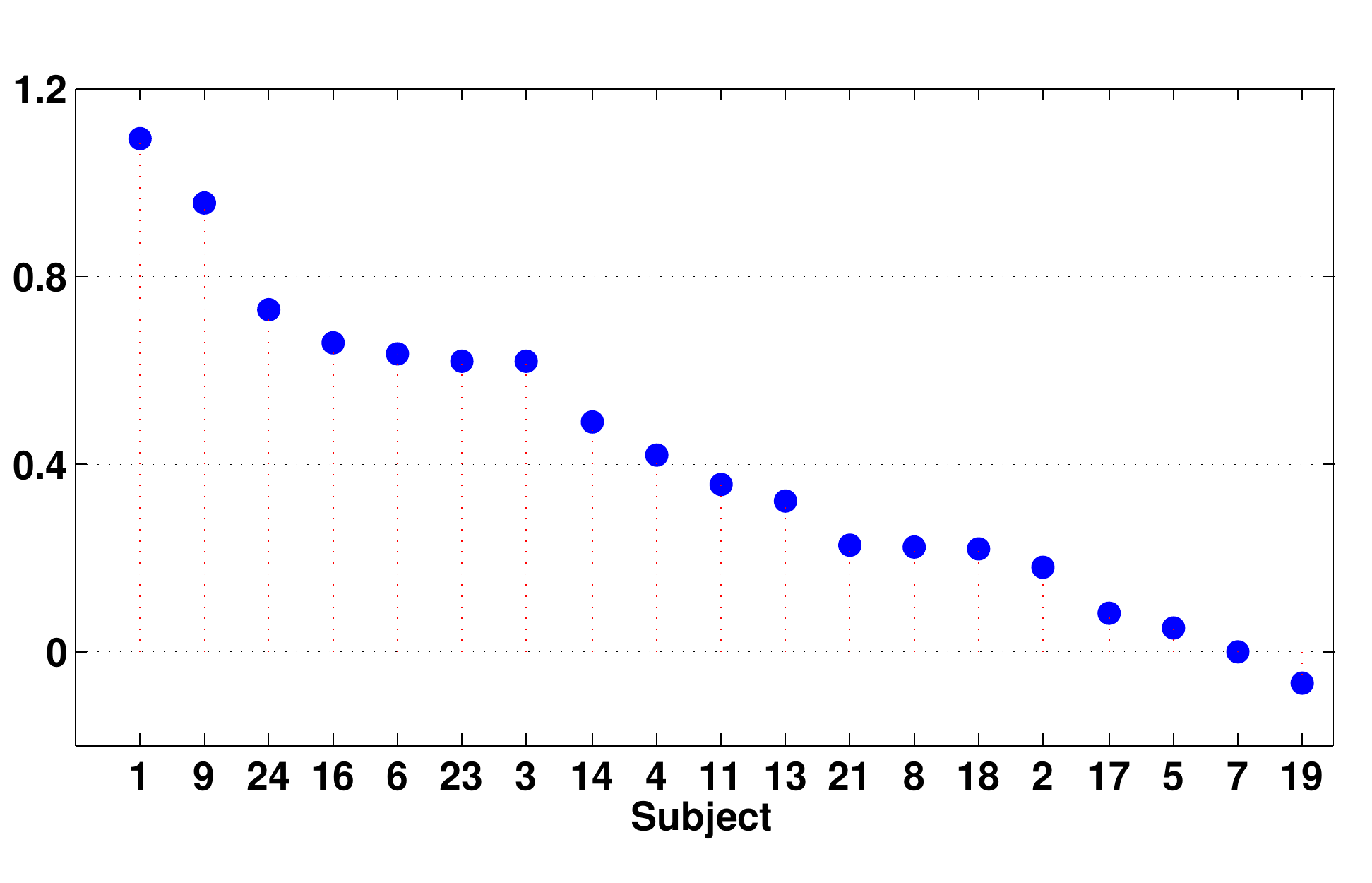}
		\vspace{-0.3cm}
    \caption{Risk attitudes of $19$ subjects (indexed on $x$-axis) derived by the (\ref{eq:logi}). \label{fig:riskatt}}
    \end{center}
\end{figure}

\begin{figure}[!h]
\begin{center}
    \includegraphics[width=0.65\textwidth]{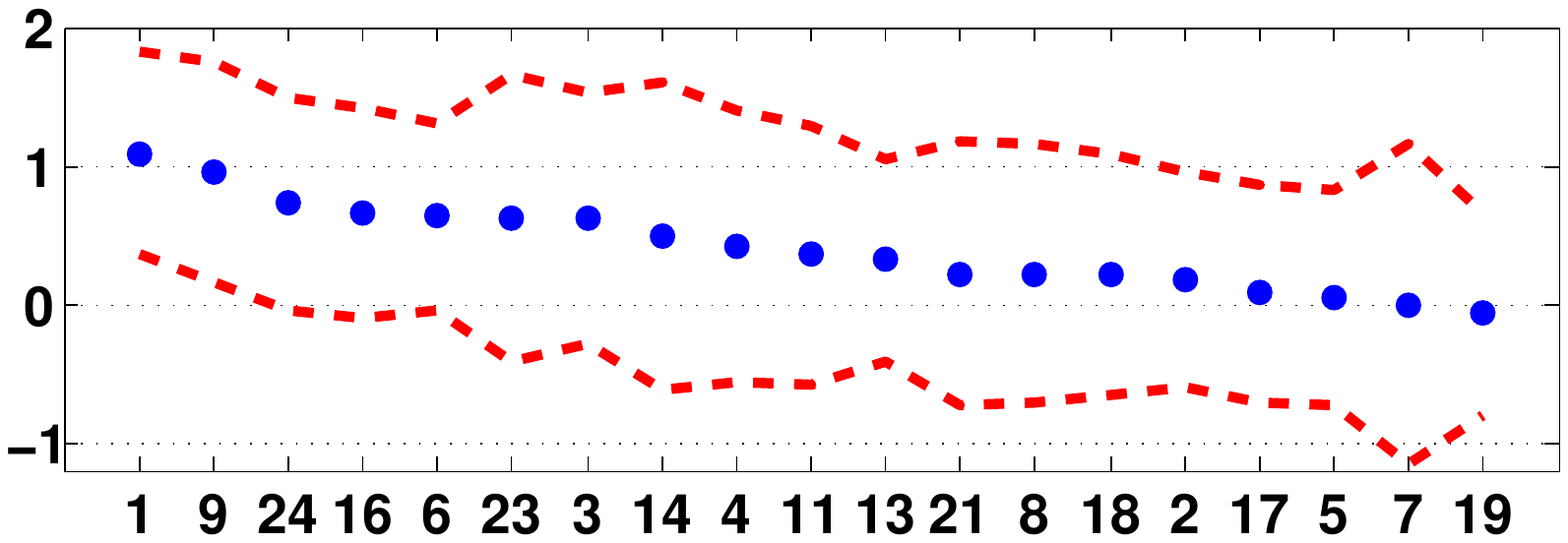}
        \caption{Sensitivity analysis of the risk attitude $\phi$: estimates $\widehat{\phi}_i, i=1,\ldots,19$ with $95\%$ confidence intervals.}
				\label{fig:risksens}
				\end{center}
\end{figure}

\begin{figure}[!h]
\begin{center}
        \includegraphics[width=0.75\textwidth]{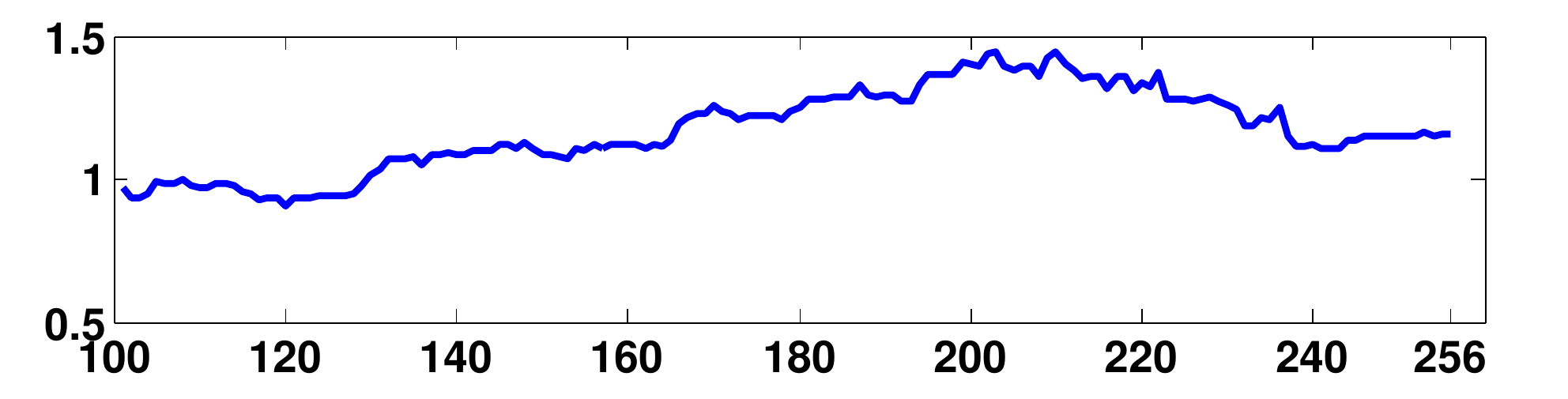}
        \caption{The derived risk attitude of subject $1$ in a rolling window exercise ($\widehat{\phi}_i$ estimated from past $100$ ID answers).}
				\label{fig:risksens1}
				\end{center}
\end{figure}

The risk attitude can be estimated based on subject responses (risky choice vs. sure, $5\%$ return) by the logistic model:
\begin{eqnarray}
\displaystyle
\Prob\left\{\text{risky choice}|x\right\}&=&\frac{1}{1+\exp\left\{\overline{x}-\phi S(x)-5\right\}}.
\label{eq:logi}
\end{eqnarray}
 Negative values of $\widehat{\phi}_i$ indicate a risk seeking behavior, $\widehat{\phi}_i\approx 0$ relates to risk-neutrality and $\widehat{\phi}_i >0$ to risk aversion. The estimated risk attitudes are shown in Figure \ref{fig:riskatt} and additional analysis in Figures \ref{fig:risksens} and \ref{fig:risksens1}. For simplicity of presentation, in the subsequent part of the analysis we show data for two most extreme subjects: $19$-th, risk-seeking: risk weight=$-0.0699$ and $1$-st, risk-averse: risk weight=$1.092$.
\section{Results}\label{results}
Choice of the model parameters is described and the clustering results together with the estimated factor loadings are presented. This $2$-step dimension reduction technique simplifies the brain dynamics into $C-$dimensional time series. The activated clusters are selected in the Activation-step and the subjects' risk aversion is modeled and predicted based only on the fMRI data. 
\subsection{Model Parameters}
Selection of the number of clusters plays of course a role in our analysis. Choosing only few regions of interest (i.e., $50$ parellations) leads to over-generalized and condensed regions that are anatomically distinctive, see, e.g., \citet{Craddock:2012}. Increasing the division into $200$ clusters is reflecting the anatomical brain atlases \citep{Talairach, Oxford} and an approach based on the brain identified atlas zones is often used. When a more precise parcellation is called for, practitioners then select $1000$ clusters as discussed by \citet{Craddock:2012}. Our study aims to find activated brain regions related to the investment decisions, where the possible HRF may be subtle. Moreover, a successful implementation of the dynamic semiparametric factor model and conducted testing procedure requires highly accurate and homogenous inputs, we thus select $C=1000$ clusters and ensure thereby the high accuracy of the representation. In the next step each (homogenous) cluster is represented by the DSFM technique with $1$ dynamic factor, $L=1$ for all cluster $c=1,\ldots,1000$. Inclusion of higher number, though yielding a better fit, does not allow for a simple interpretation.  

\begin{figure}[!h]
\begin{center}
  \includegraphics[width=0.95\textwidth]{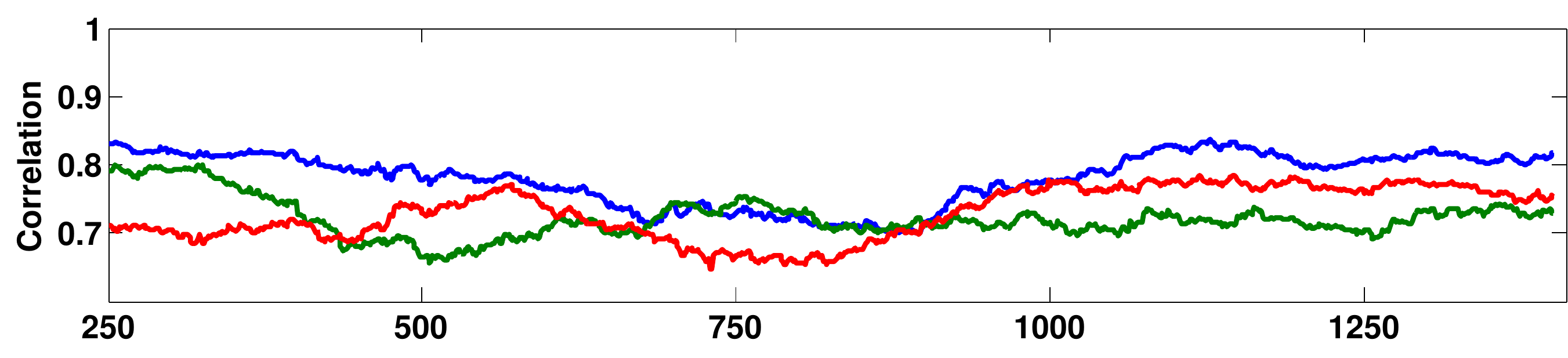} 
	\includegraphics[width=0.95\textwidth]{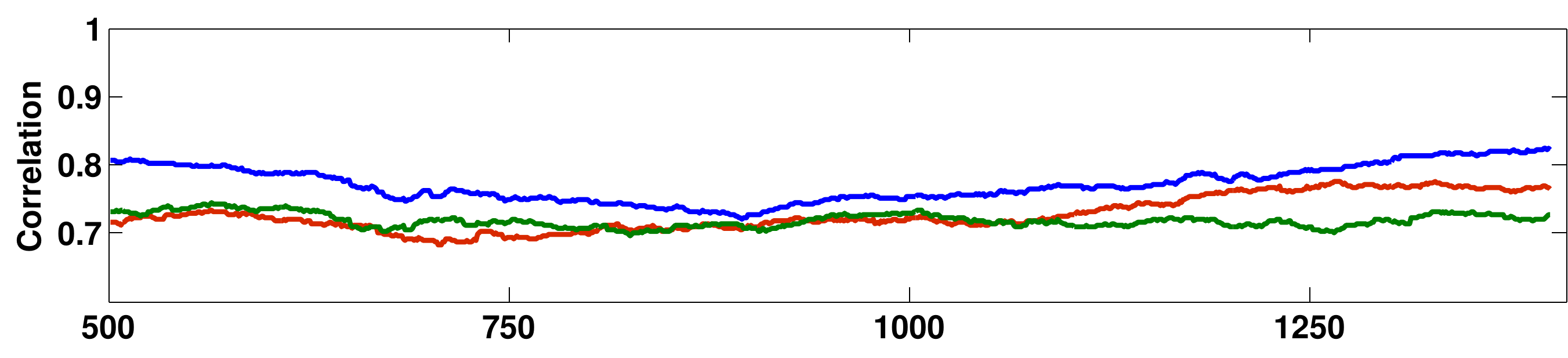}
\end{center}
 \caption{Time series of the correlation coefficient derived by the rolling
window (250 top, 500 bottom) for the center voxel and: horizontal, vertical
diagonal neighboring voxel for aINS(right) of subject 1.}
 \label{fig:corry}
\end{figure}

The parcellation technique is based on (\ref{eq:similarity}) as a proximity measure. In order to check stability of (\ref{eq:similarity}) over the entire experiment we conduct a moving window exercise. Figure \ref{fig:corry} shows the correlation between $3$ neighboring voxels derived by a rolling window exercise (for past $250\approx 8$ min and $500\approx 17$ min). One observes a stable, stationary behavior over time which stands in favor of our modelling setup.   
\subsection{Clustering Results}\label{clustering results}
\begin{figure}[!h]
\begin{center}
\begin{tabular}{ccc}
  \includegraphics[height=3.5cm]{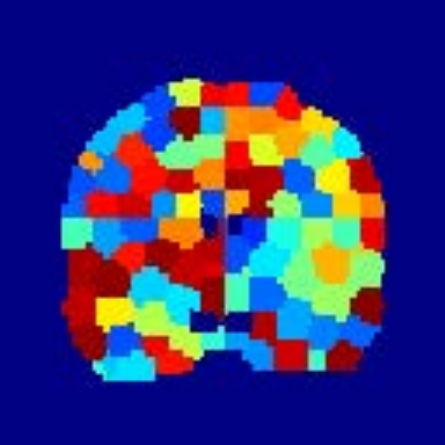}
&
   \includegraphics[height=3.5cm]{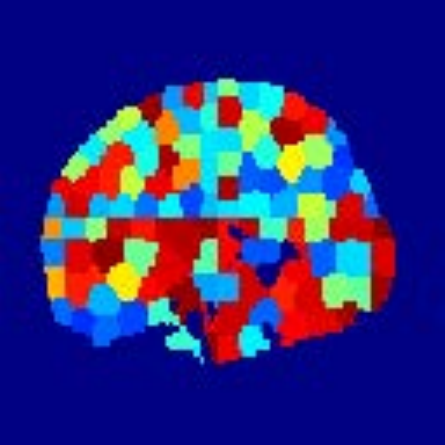}
   &
   \includegraphics[height=3.5cm]{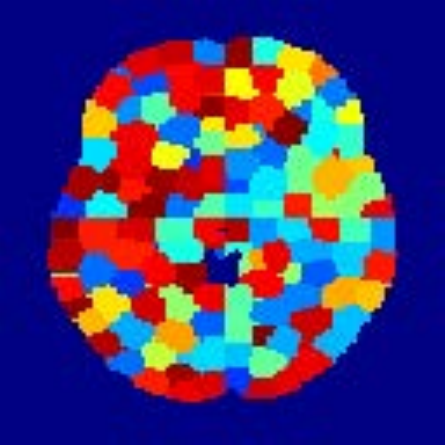}
\\

   \end{tabular}
\end{center}
 \caption{Illustration of the clustering results for subject $1$ derived by the NCUT algorithm, $C=1000$. The parcellation is represented as an orthogonal view and color-coding is arbitrarily used to capture the clusters' boundaries.}
 \label{fig:Clust}
\end{figure}

\begin{figure}[!h]
\begin{center}
\begin{tabular}{cc}
  \includegraphics[width=0.5\textwidth]{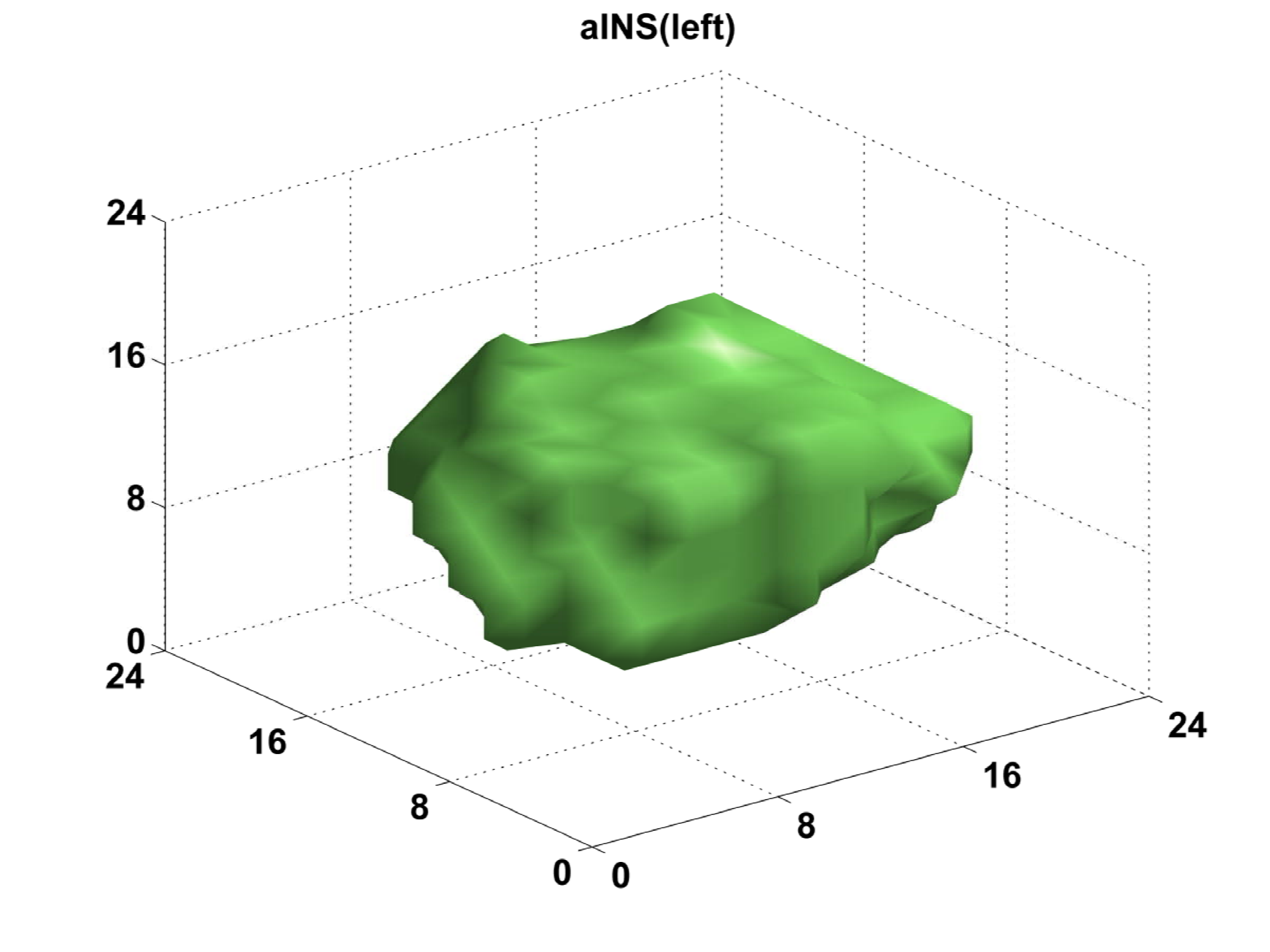}&\includegraphics[width=0.5\textwidth]{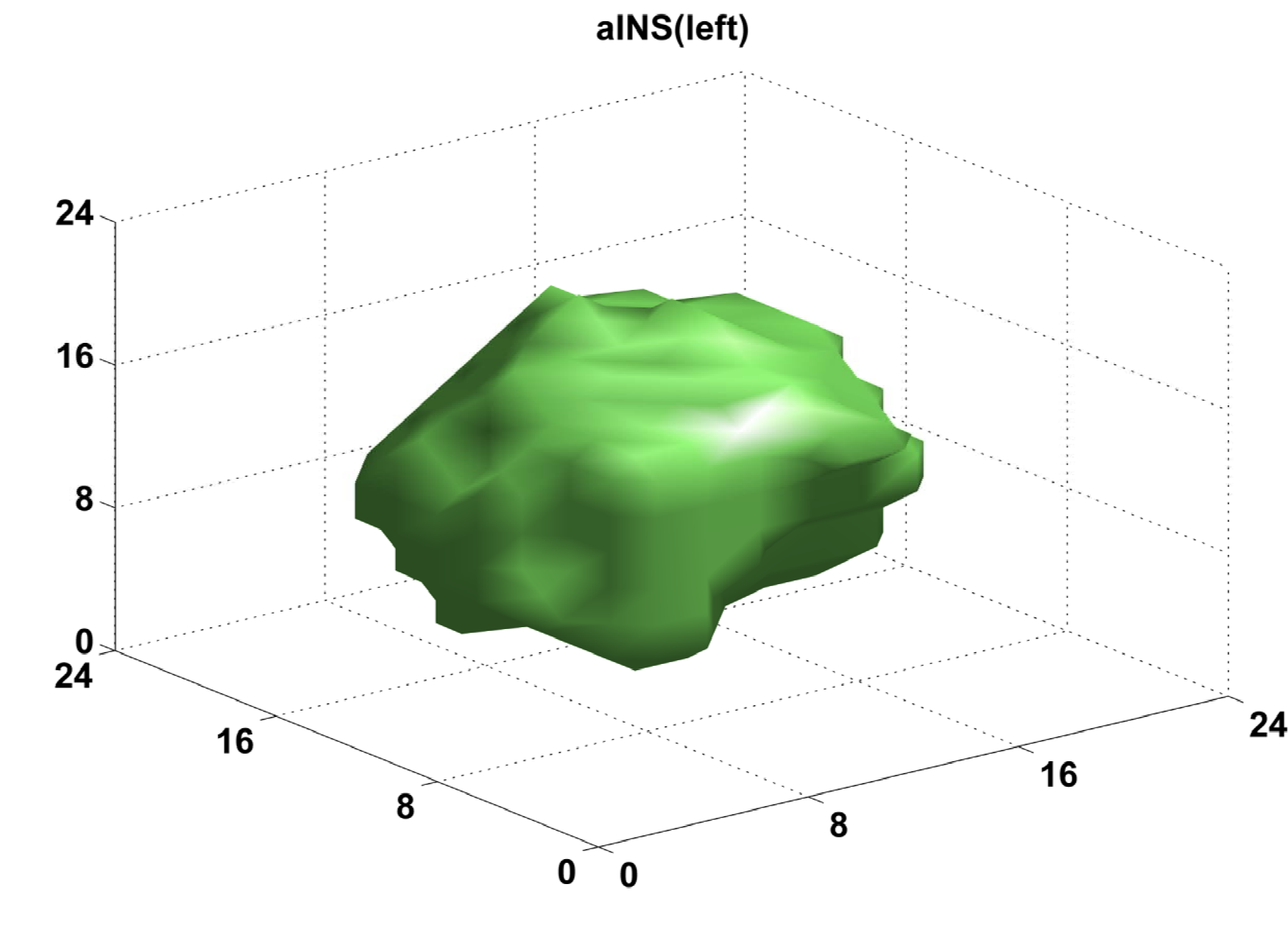}\\
	  \includegraphics[width=0.5\textwidth]{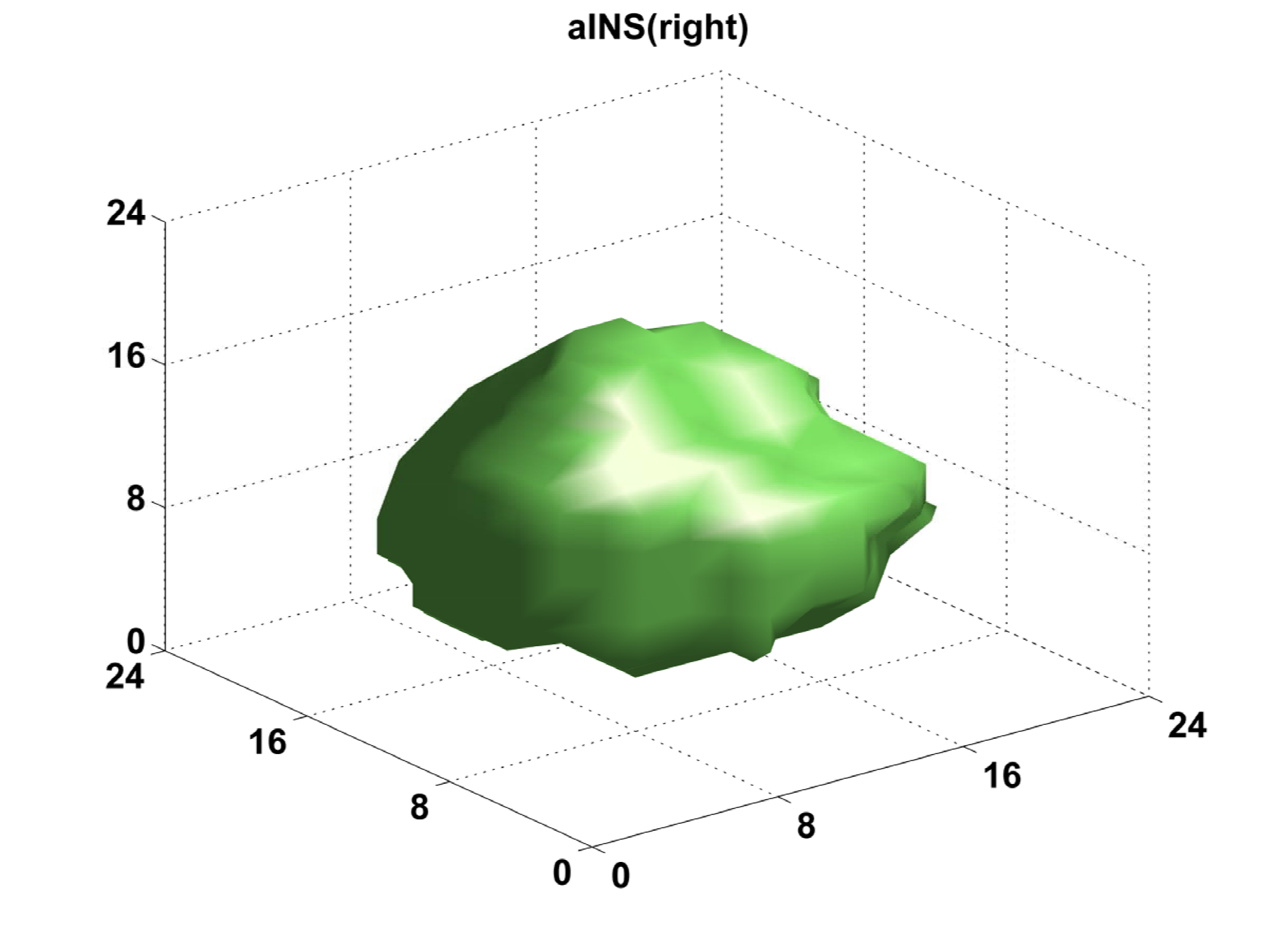}&\includegraphics[width=0.5\textwidth]{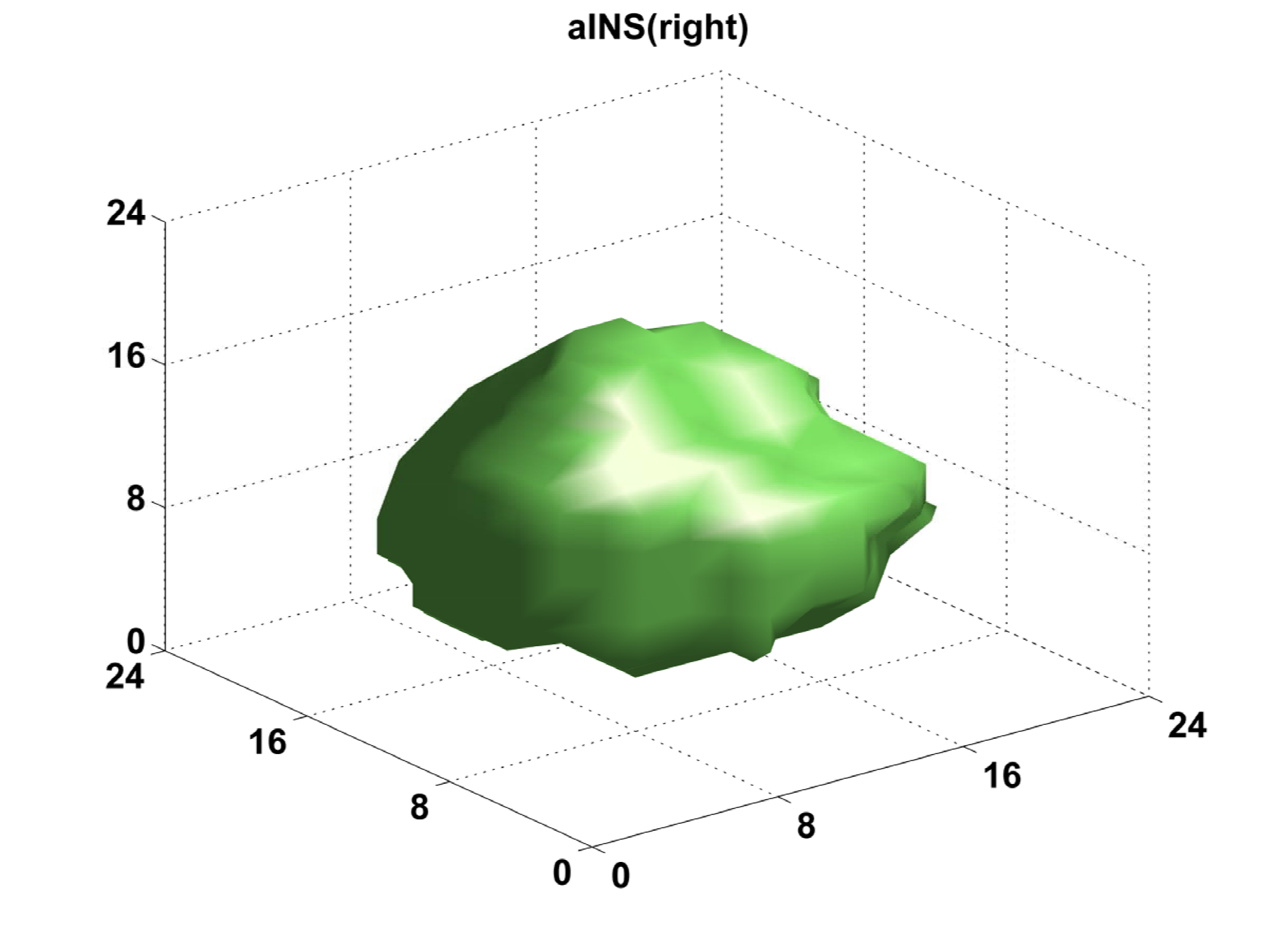}\\
    \includegraphics[width=0.5\textwidth]{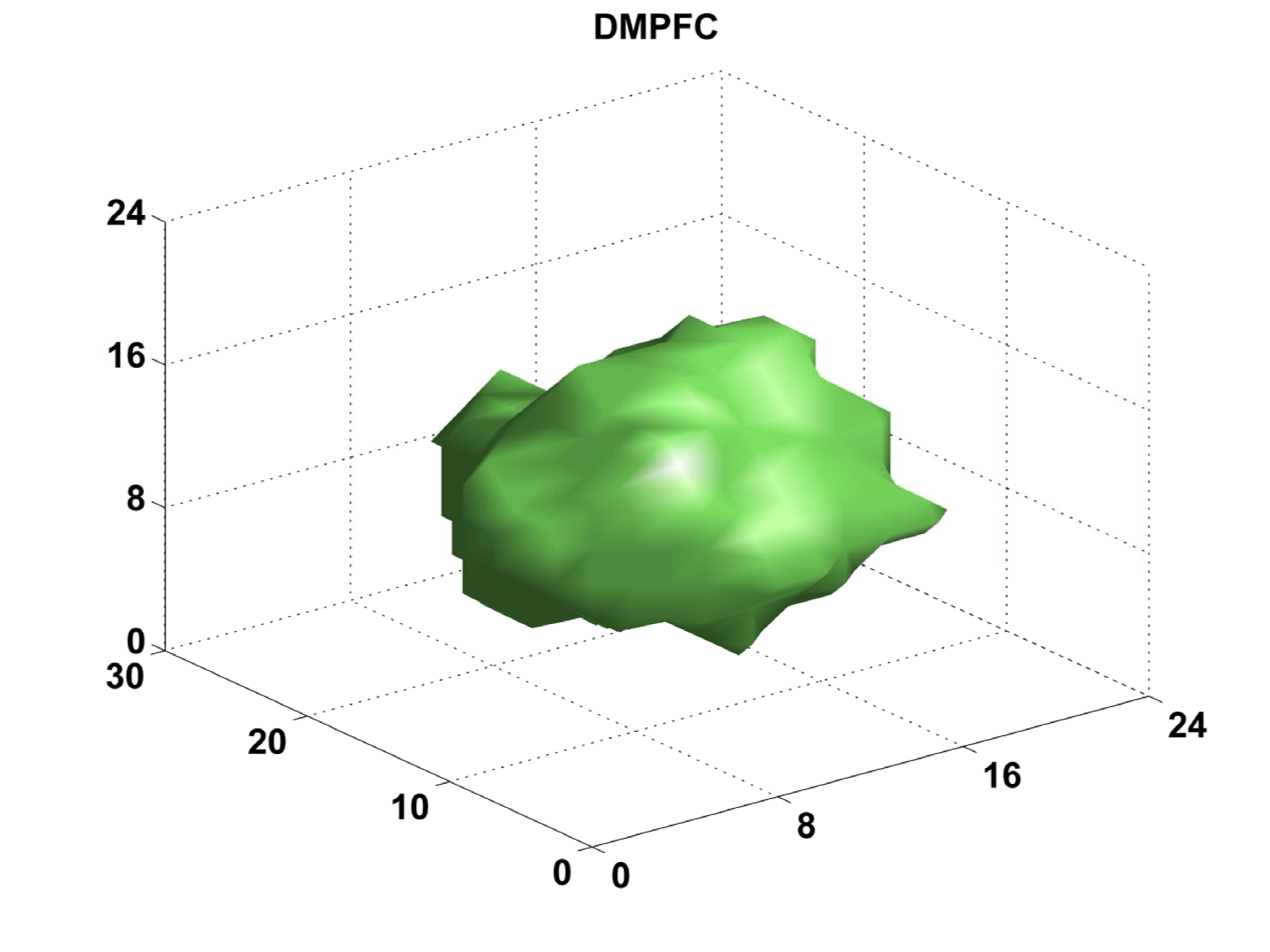}&\includegraphics[width=0.5\textwidth]{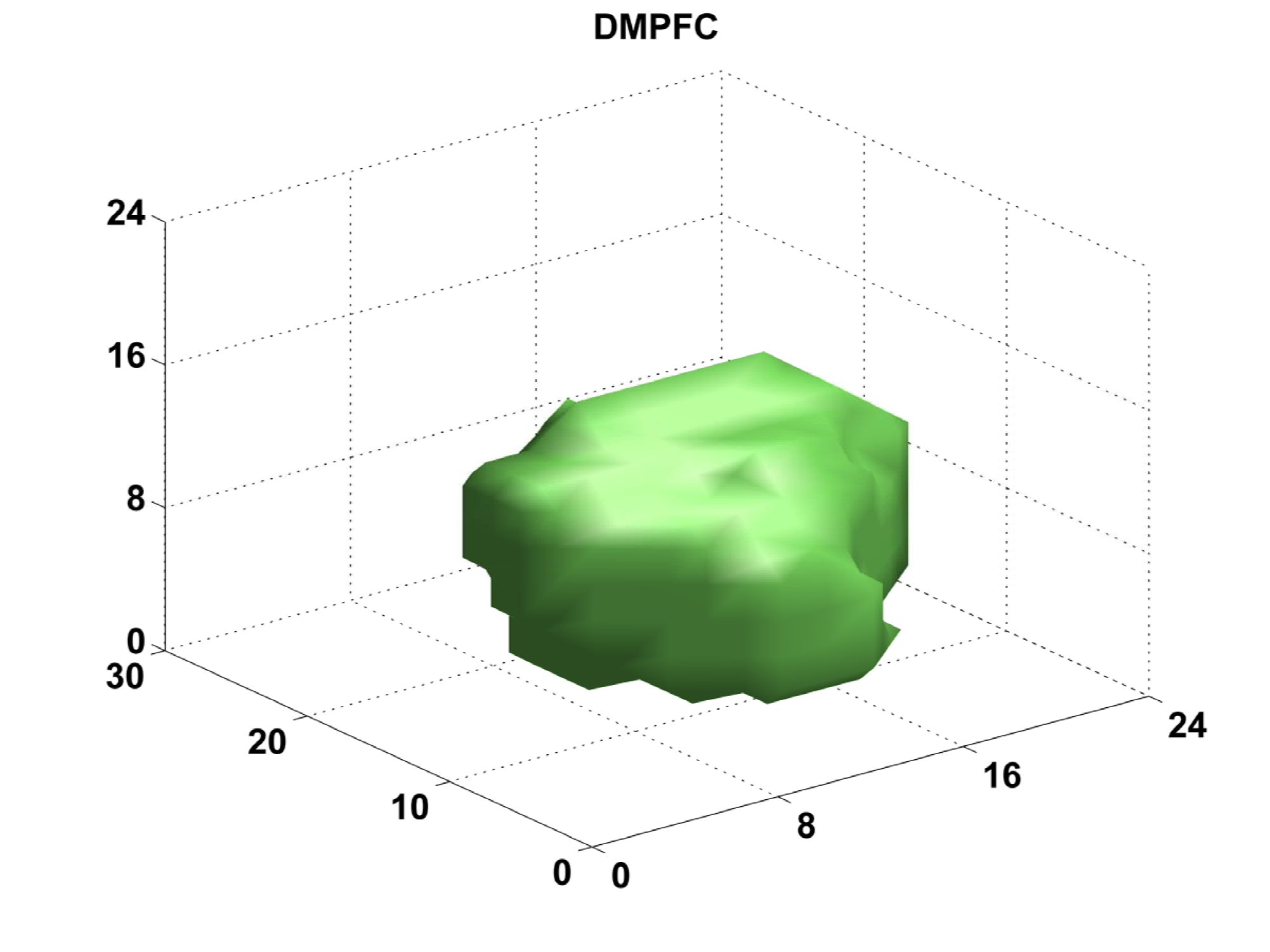}
   \end{tabular}
\end{center}
 \caption{Contour plots of derived  aINS(left), aINS(right) and DMPFC (upper, middle lower panel) clusters for subjects $1$ (left) and $19$ (right), respectively; derived by the NCUT algorithm with $C=1000$. $x$-, $y$- $z$-axis denote the $3D$ space given in millimeters.}
 \label{fig:Selclust}
\end{figure}

Clustering results are illustrated in Figure \ref{fig:Clust}. The subject-specific parcellation, though computationally extensive, addresses inter-subject functional variability. Therefore, we derive spatially coherent regions of homogenous functional connectivity, that are present at a voxel scale. The clusters are contiguous sets of neighboring voxels and a distinction between network nodes and large-scale network of nodes is ensured, see \citet{smith2009}.  The neuroscientific interpretability is preserved and further elaborated on in the modelling and testing part of our study. An average cluster is of a size $207$ voxels, which might be compared to a $6\times6\times6=216$ ($12$ mm) cube. The smallest cluster is a singleton and the largest consist of $353$ voxels. Clusters have a data-driven shape and vary with respect to the size and spatial structure as shown in Figure \ref{fig:Selclust}. 

\subsection{Factor Loadings $\widehat{Z}_{t}$}\label{loadings}

\begin{figure}[!t]
\begin{center}
aINS(left)\\
\includegraphics[height=7.4em]{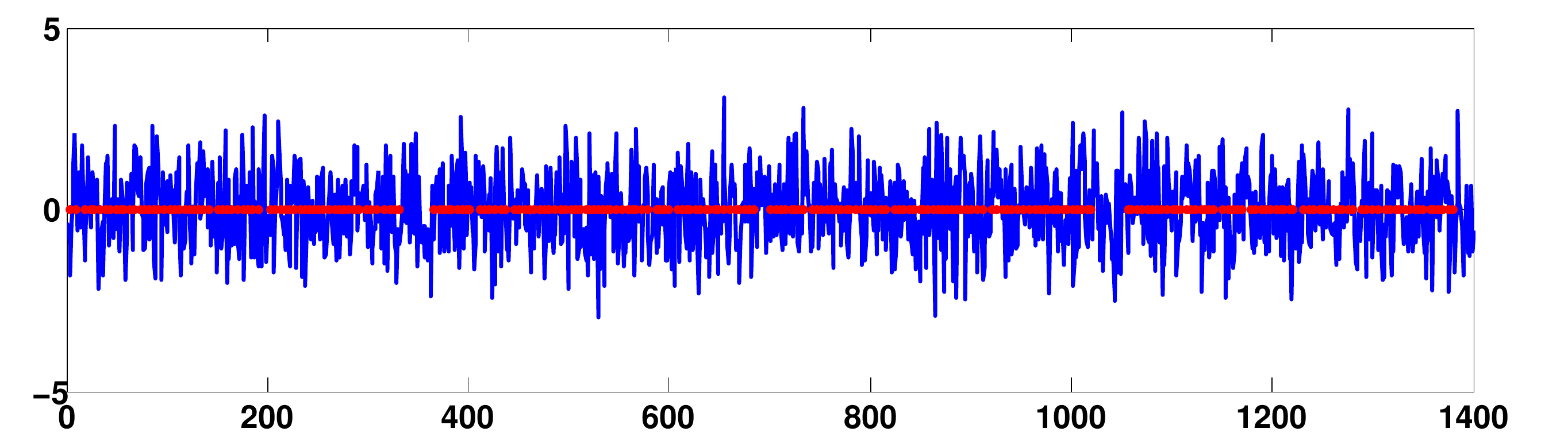}\\ 
\includegraphics[height=7.4em]{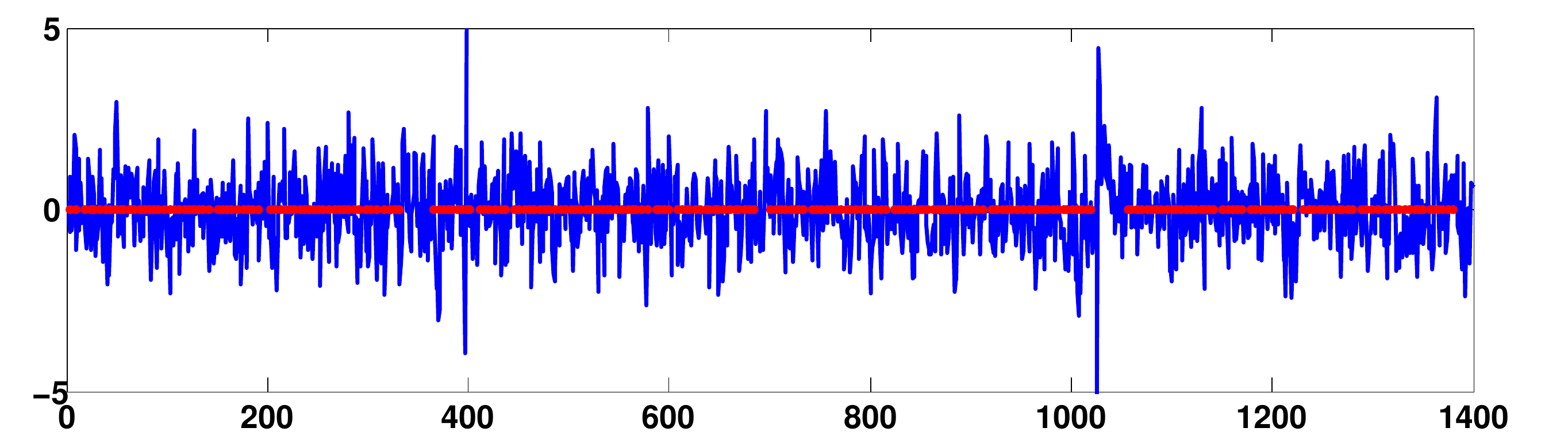}\\
aINS(right)\\
\includegraphics[height=7.4em]{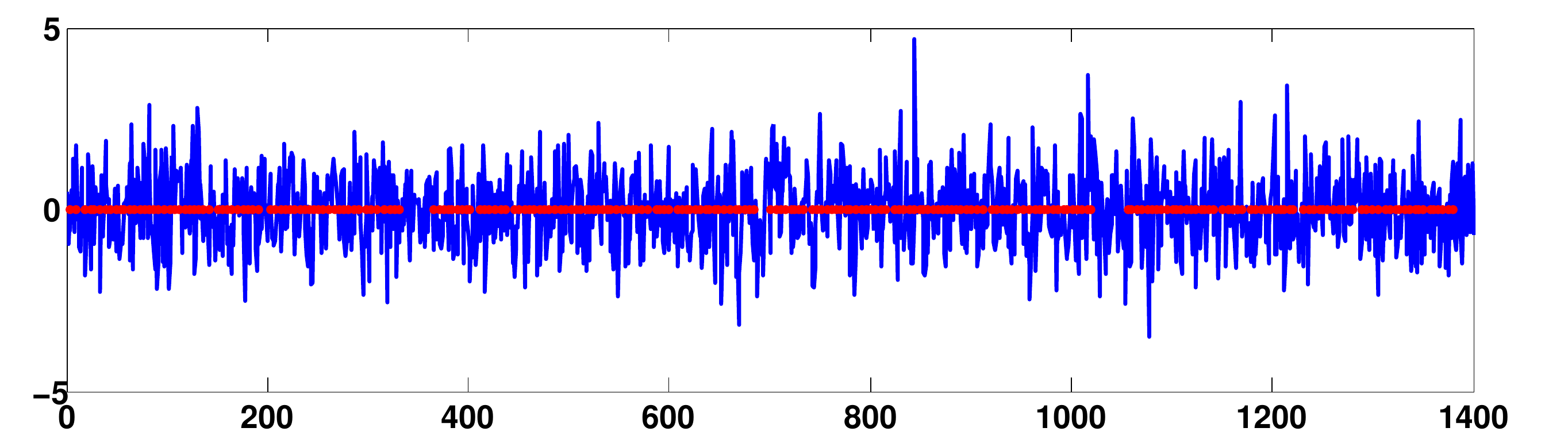}\\
\includegraphics[height=7.4em]{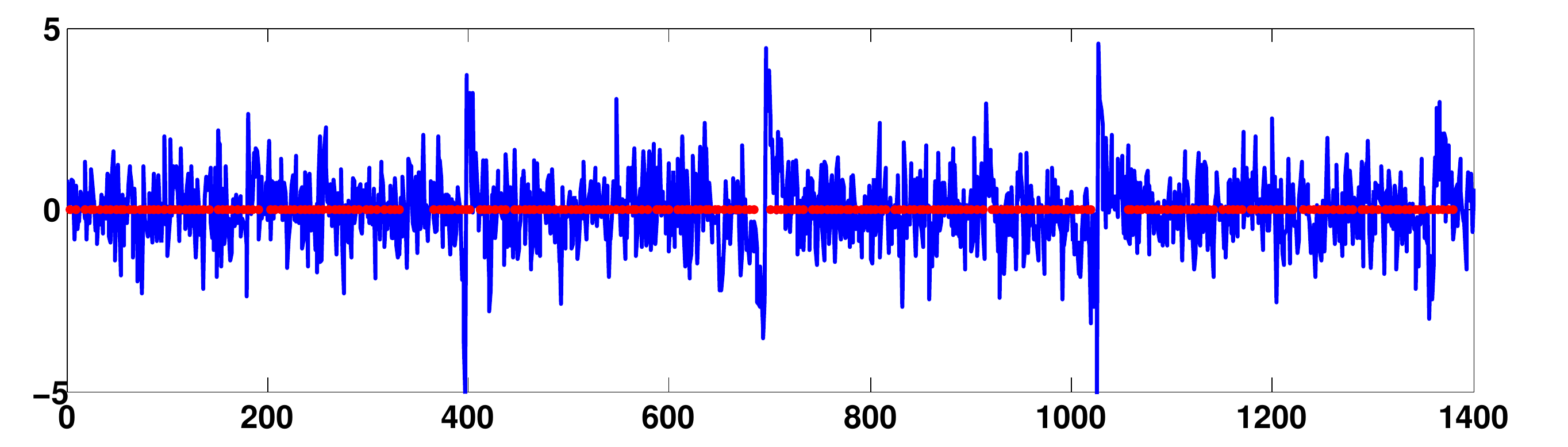}\\
DMPFC\\
\includegraphics[height=7.4em]{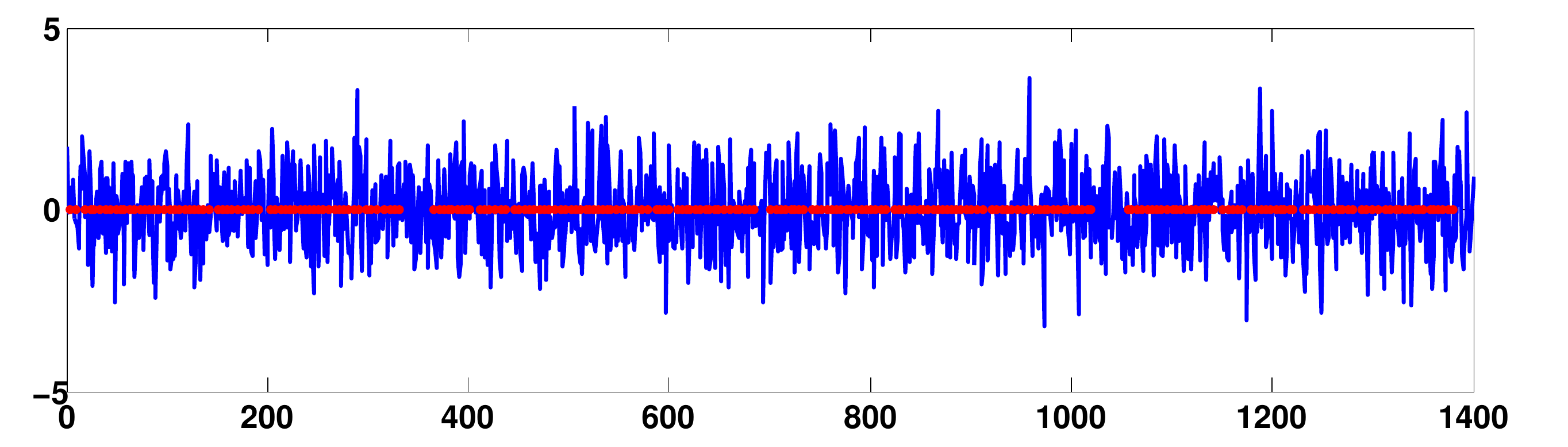}\\
\includegraphics[height=7.4em]{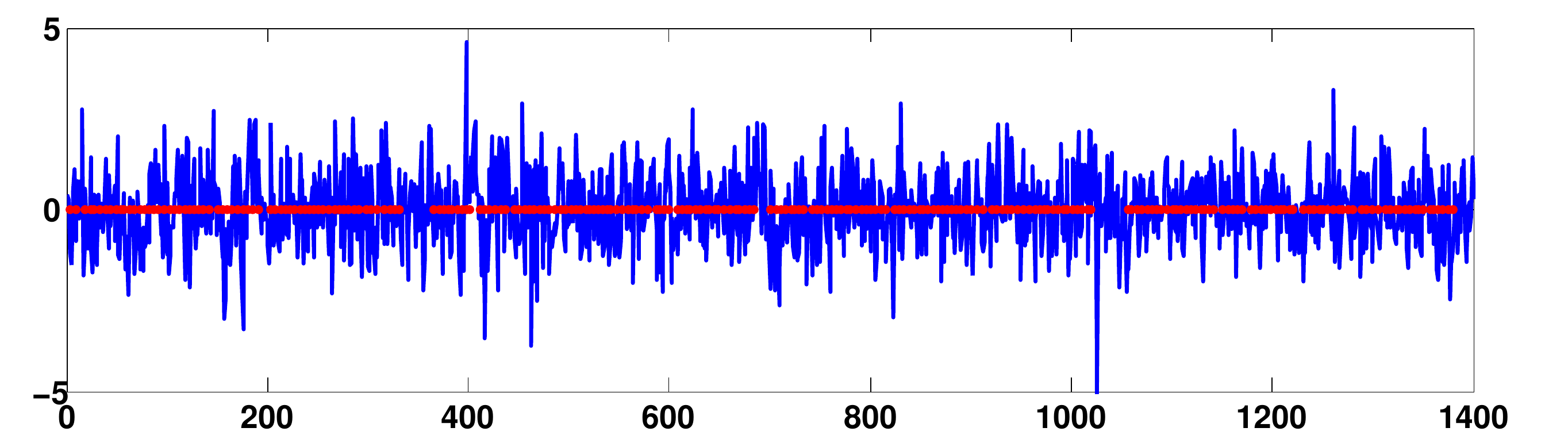}\\
\end{center}
\vspace{-0.2cm}
\caption{Factors loadings $\widehat{Z}_{t}$ for clusters aINS(left), aINS(right) and DMPFC (upper, middle lower panel) for risk averse subject $1$ (top)
and weakly risk seeking subject $19$ (bottom) plotted against time (each $2$ seconds). Red points correspond to the time points of stimuli.}
\label{fig:zets}
\end{figure}

\begin{figure}[!h]
\begin{center}
\begin{tabular}{cc}
\includegraphics[width=0.5\textwidth]{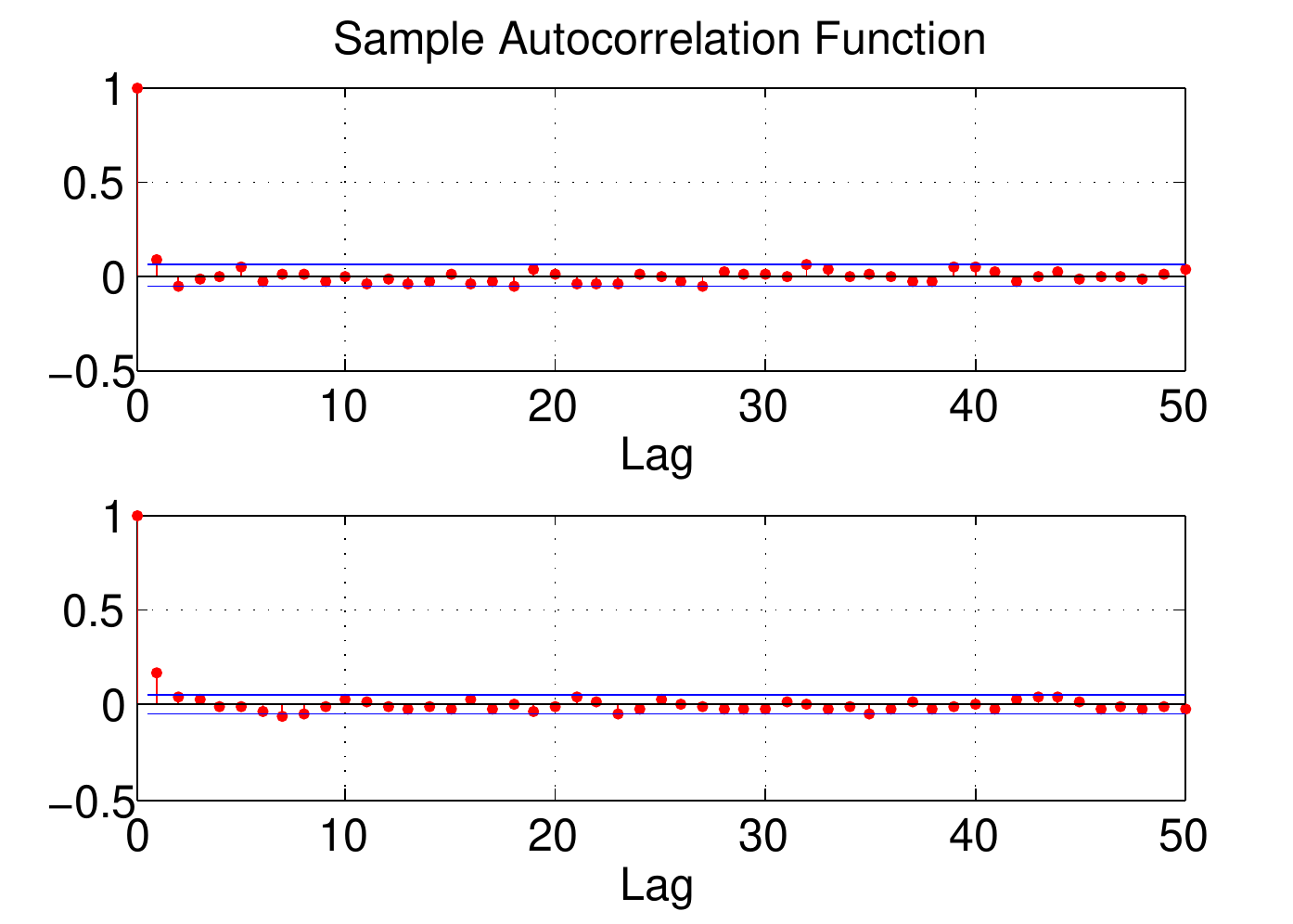}&
\includegraphics[width=0.5\textwidth]{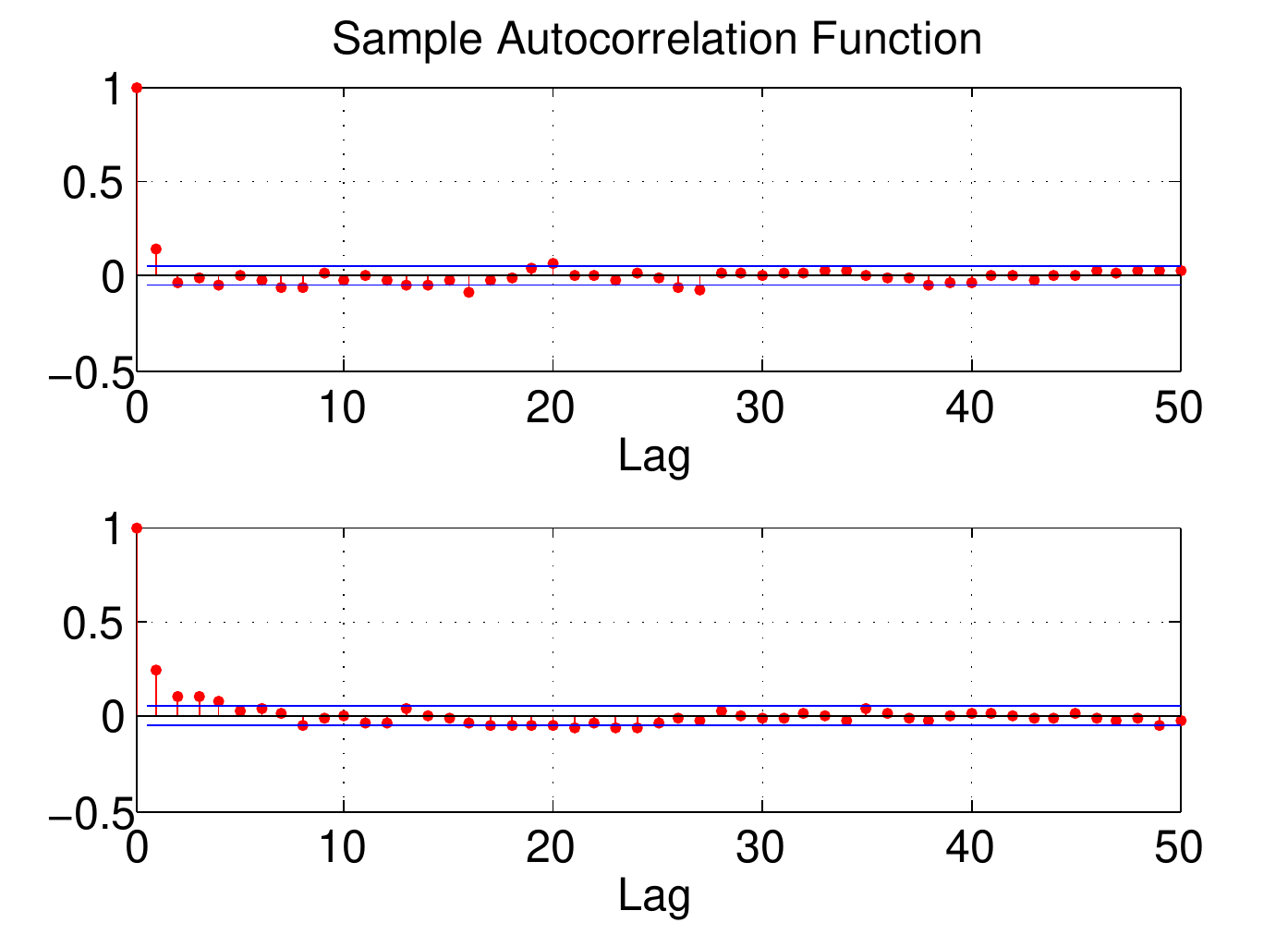}\\
 \multicolumn{2}{c}{\includegraphics[width=0.5\textwidth]{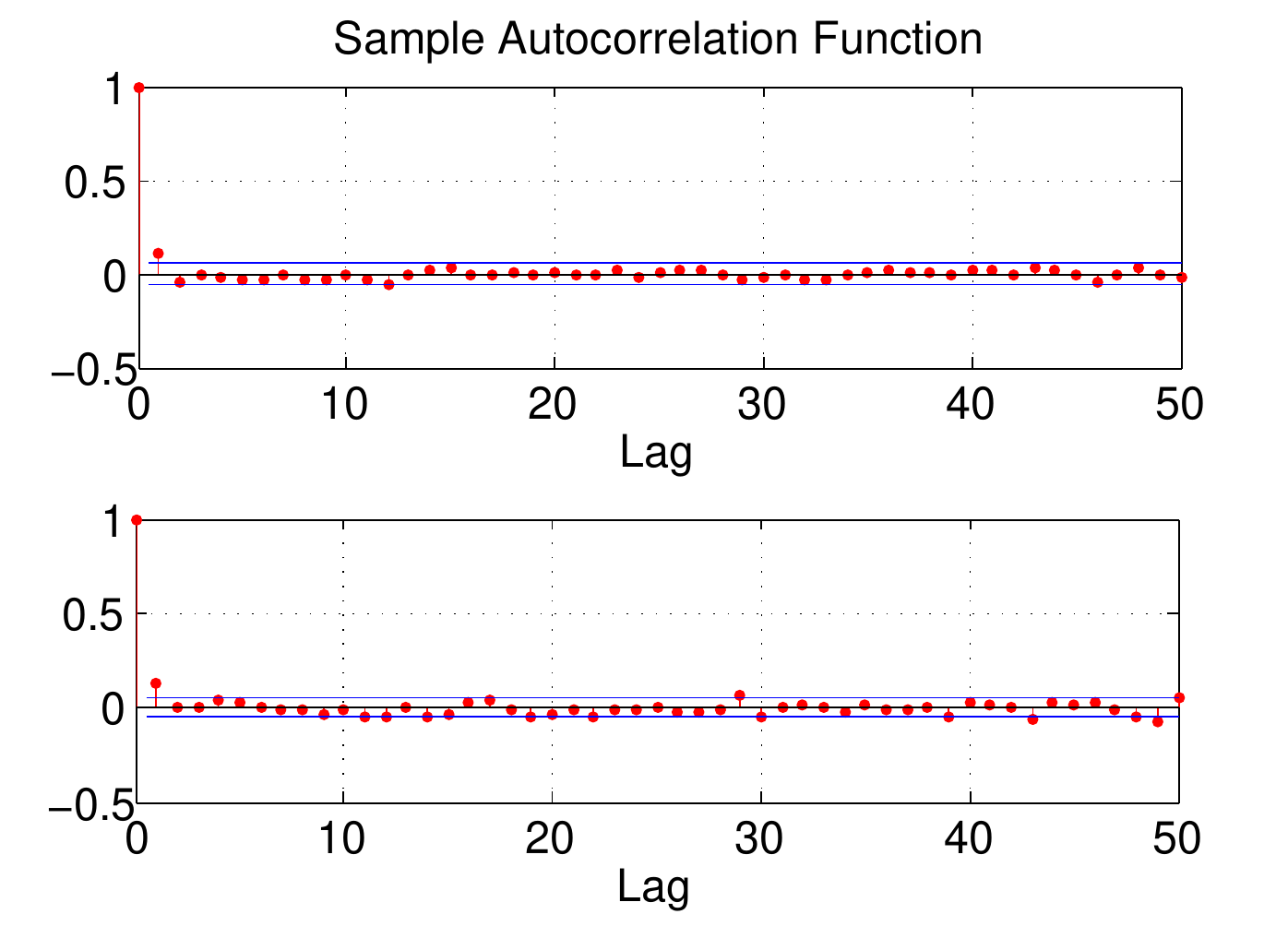}}
\end{tabular}
\end{center}
\caption{Sample autocorrelation function of aINS(left), aINS(right) and DMPFC $\widehat{Z}_t$ (top left, top right, bottom panel, respectively)  for subjects $1$ (top) and $19$ (bottom), respectively.}
\label{fig:acf}
\end{figure}

The clustering spatial maps serve as a basis for further exploratory analysis. The information carried in time evolution of the derived clusters is extracted by the DSFM technique. More precisely, all voxels belonging to cluster $c$ of subject $i$: $Y_{c,1}^{i},\ldots,Y_{c,J_c^{i}}^{i}$, where $J_c^{i}$ is the size of $c$ cluster for subject $i$, are jointly modeled by (\ref{eq:DSFM}). For simplicity of representation and as a natural consequence of cluster (homogenous) construction we employ the DSFM with $L=1$. Thus, each cluster's dynamics are captured by the univariate time series $\widehat{Z}_{t}^{i,c}$, $i=1,\ldots, I$; $c=1,\ldots,1000$, and the complete brain representation consist of $1000$ processes. The derived brain model significantly simplifies the complexity of the data, while ensuring the interpretability and a good quality fit. For a demonstration two extreme subjects: 1 (with the smallest risk attitude) and 19 (with the largest risk attitude) are selected, see Figure \ref{fig:riskatt}. Figure \ref{fig:zets} shows the estimated
$\widehat{Z}_{t}^{1}$ and $\widehat{Z}_{t}^{19}$ for anterior insula (aINS; left and right) and dorsomedial prefrontal cortex (DMPFC) clusters. All factor loadings exhibits stationary behavior, high persitency and a high fluctuation around their mean value (see Figure \ref{fig:acf} and Table \ref{tab:stationarity}), which may be related to the underlying investment decision stimulus.  

\subsection{Activation Results $\widehat{Z}_{t}$}\label{activation}
The derived low-dimensional representation of each cluster $\widehat{Z}_{t}$  serves as a principal unit of this study and is tested for activation. We compare our method with both, the standard voxelwise GLM technique and the approach proposed by \cite{Heller} (average over voxels and use it as a cluster temporal representation). Four separate analyses were conducted (single, correlated and uncorrelated, jointly all types of portfolio). For each type of investment we reported the same activation pattern, thus only the joint analysis (all portfolios) is reported here. 
\begin{figure}[h!]
\begin{center}
\begin{tabular}{ccc}
  \includegraphics[height=3.5cm]{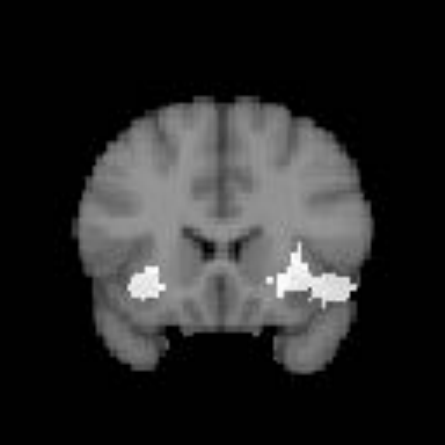}&\includegraphics[height=3.5cm]{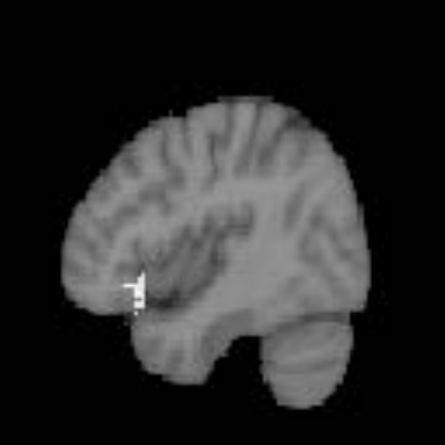}&\includegraphics[height=3.5cm]{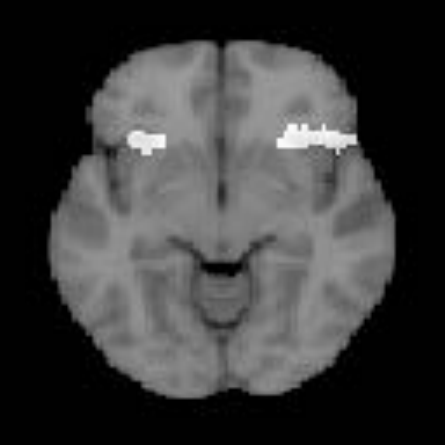}\\
\includegraphics[height=3.5cm]{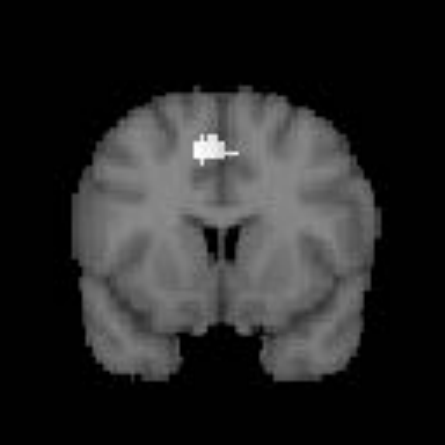}&\includegraphics[height=3.5cm]{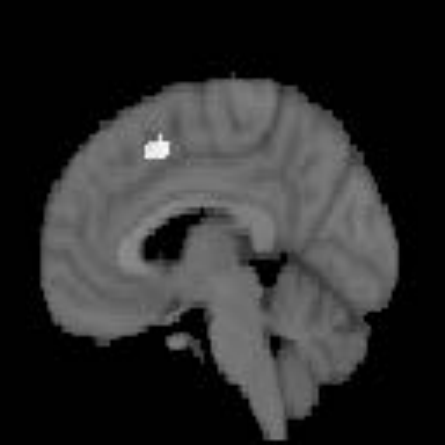}&\includegraphics[height=3.5cm]{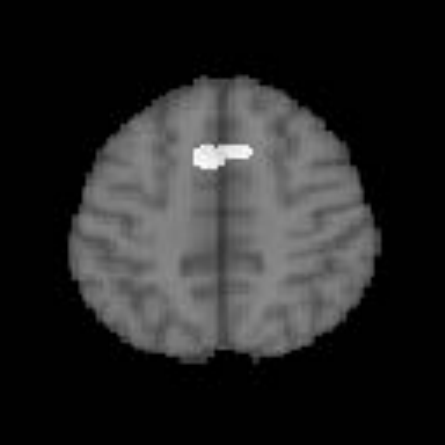}\\
   \end{tabular}
\end{center}
\vspace{-0.3cm}
 \caption{Results of the higher-level analysis (mixed-effects model) associated with decision making; $Z$-scores$>3.09$. Upper panel: the bilateral aINS, lower panel: DMPFC. }
 \label{fig:Actnii}
\end{figure}

\begin{figure}[!h]
\begin{center}
\begin{tabular}{ccc}
  \includegraphics[width=0.32\textwidth]{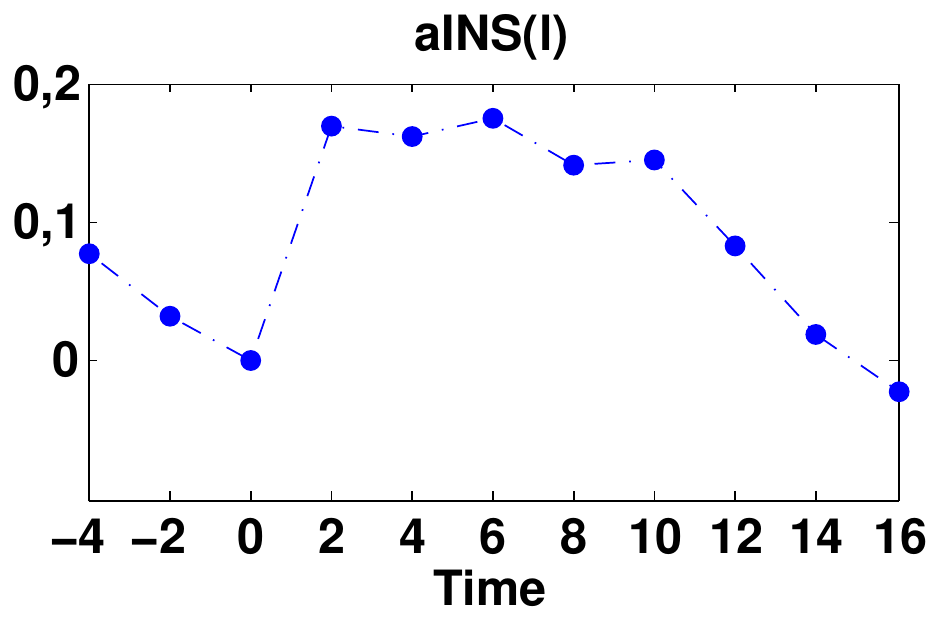} &
	\includegraphics[width=0.32\textwidth]{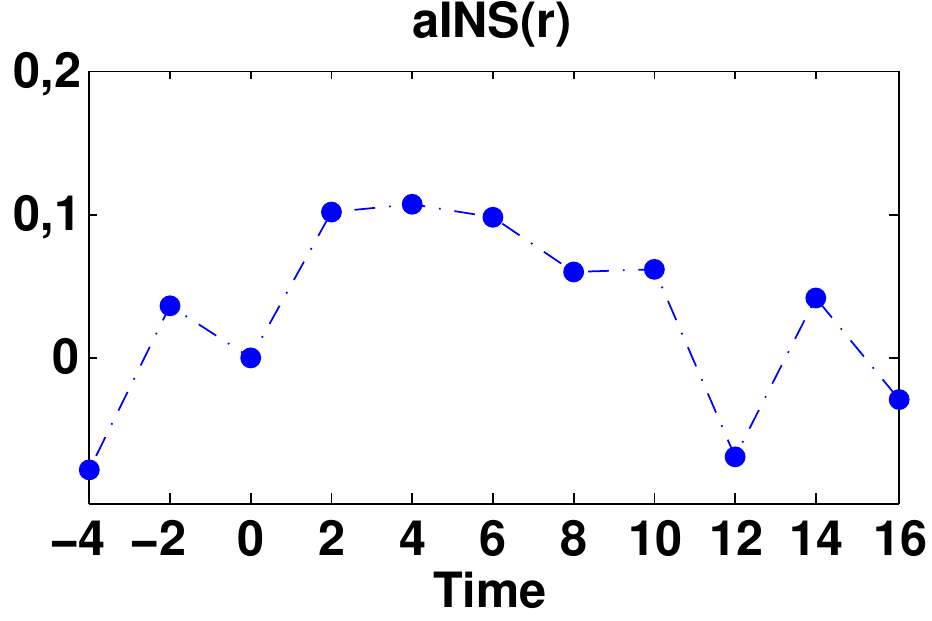} &
		\includegraphics[width=0.32\textwidth]{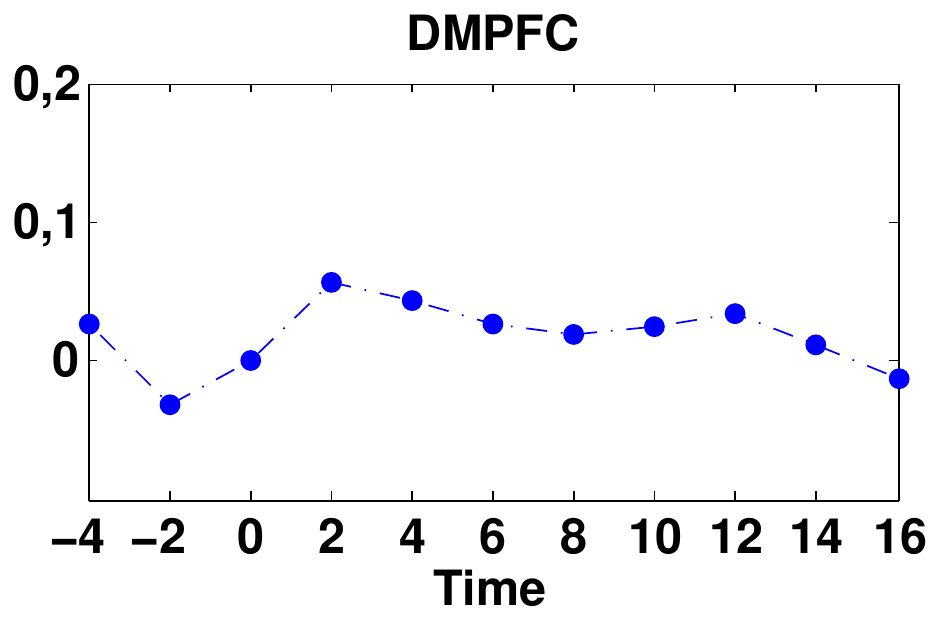}
		\end{tabular}
\end{center}
 \caption{Average reaction to the ID stimulus over all $19$ subjects for bilateral aINS and DMPFC regions plotted against time (from -$4$ seconds before the stimuli up to $16$ seconds afterwards).}
 \label{fig:HH}
\end{figure}

Figure \ref{fig:Actnii} presents significant brain correlates of the ID task: aINS and DMPFC associated with decision making. These activation results are in line with findings by \cite{mohr:2010, mohr:2010b} and contribute to the neural foundations of risk-return model. Altogether $9$ activated clusters were detected which survived statistical thresholding at $Z$-scores$>3.09$ and had a cluster size of at least $20$ voxels. Besides aINS and DMPFC factors corresponding to decision making, we identified other brain regions previously associated with visual perception and motoric responses. These factors are most likely not connected to the decision making process but confirm the activity of regions which were necessary to give the answer by pushing the button. Average reactions to the ID stimuli over all $19$ subjects are depicted in Figure \ref{fig:HH}. Reported maximum $Z$-scores for aINS and DMPFC are shown in Table \ref{tab:zscore}. One observes that all approaches yield very similar results, though the highest maximum $Z$-score is achieved by the GLM technique for all $3$ ROIs. Secondly, the DSFM outperforms the simple averaging over voxels. The non-parametric estimation pays off in terms of the quality of the representation. 

\section{Risk Attitude $\backslash$ Stimulus Response}
The key goal in neuroeconomics is to "(\ldots) ground economic theory in detailed neural mechanisms which are expressed mathematically and make behavioral predictions." as \cite{Camerer, Camerer2013} states. Motivated by that, we investigated a connection between the neural processes underlying decision making and risk perception. Without prior knowledge of the subjects' answers, based only on the activated cluster dynamics, represented by $\widehat{Z}_{t}$ a simple model is proposed to predict the risk attitude $\phi_i$. As described in section (\ref{activation}) three activated (see Table \ref{tab:zscore}) clusters are associated with decision making under risk. Therefore only cluster dynamics of bilateral aINS and DMPFC are considered here as regressors for the risk attitude $\phi_i$. These loadings (brain regions) respond to the stimulus and thus mimic neural processes present in a whole cluster during investment decisions under risk in our study. The hemodynamic response function usually peaks around $6$ seconds after the stimulus. Therefore, we focus on an average reaction to $r$, $r=1,\ldots,256$, stimulus for the $i$-th subject: $\Delta\widehat{Z}^i_{r}=\frac{1}{4} \sum_{\tau=1}^{4}\widehat{Z}_{r+\tau}^{i}-\widehat{Z}_{r}^{i}$. $\Delta\widehat{Z}^i_{r}$ covers a period up to $8$ seconds afterwards and ensures that the HRF maximum is captured. An average reaction to all stimuli (entire experiment) for a single cluster is defined as $\overline{\Delta}\widehat{Z}^i=\frac{1}{256}\sum^{256}_{r=1}\Delta\widehat{Z}^i_{r}$. Our model-free methodology closely follows the statistics proposed by \cite{myfmri, Brown}.

Understanding which among the variables: $\overline{\Delta}\widehat{Z}_{aINS(l)}$, $\overline{\Delta}\widehat{Z}_{aINS(r)}$, $\overline{\Delta}\widehat{Z}_{DMPFC}$  are related to the $\phi$ and an exploration of the forms of these relationships is done via regression analysis. More precisely: 
\begin{equation}
\phi_{i}=\alpha_0+\alpha_{1}\cdot \overline{\Delta}\widehat{Z}^i_{DMPFC}+\alpha_{2} \cdot \overline{\Delta}\widehat{Z}^i_{aINS(l)}+\alpha_{3}\cdot \overline{\Delta}\widehat{Z}^i_{aINS(r)}+\widetilde{\varepsilon}^{i},
\label{eq:risk}
\end{equation}
where $\alpha_0$ is an intercept, $\alpha=(\alpha_1,\alpha_2,\alpha_3)^\top$ is a vector of regression coefficients and $\widetilde{\varepsilon}$ stands for the error term. In other words, (spatially constrained, local) information extracted from the BOLD signal serves as regressors for the subject's risk weights.

\begin{figure}
\begin{center}
\begin{tabular}{cc}
  \includegraphics[width=0.48\textwidth]{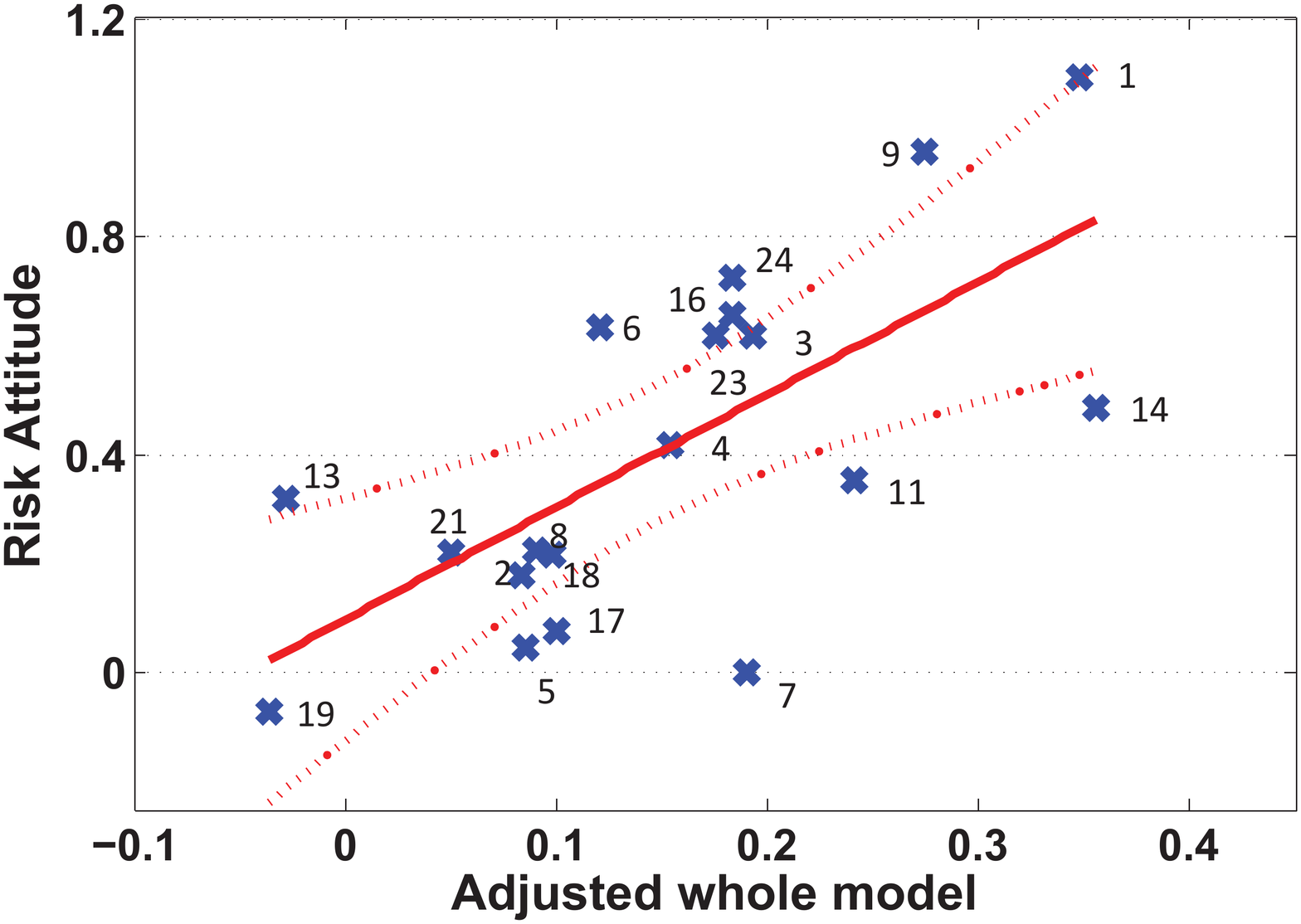}&\includegraphics[width=0.48\textwidth]{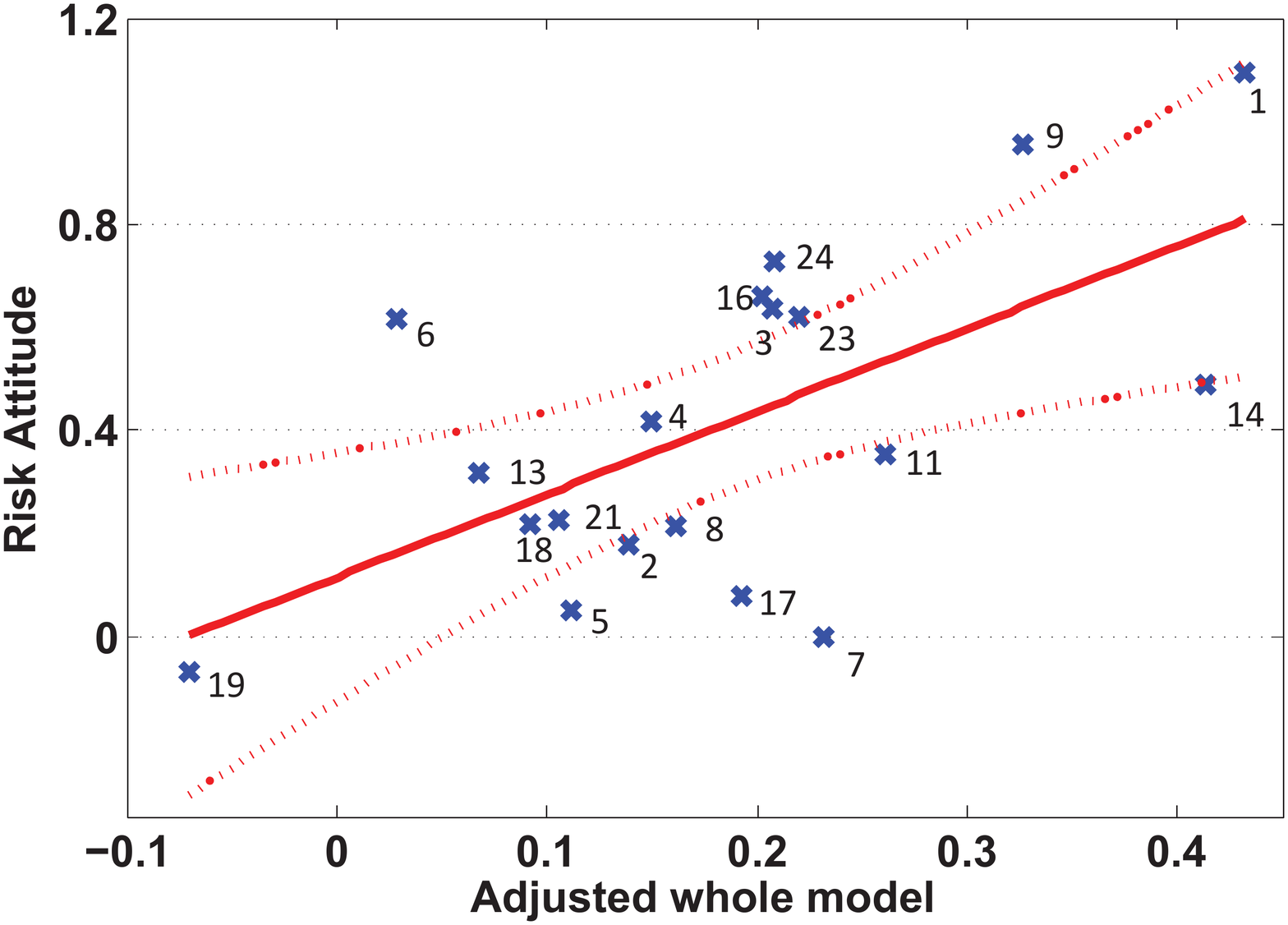}
	\end{tabular}
\end{center}
\vspace{-0.15cm}
\caption{Added variable plot for models given in (\ref{eq:risk}) left, (\ref{eq:risk1}) right panel, respectively. Horizontal axis denotes the (rescaled) best linear combination of regressors $\overline{\Delta}\widehat{Z}$ that fit $\phi$.}
\label{fig:regs}
\end{figure}

\begin{table}[h]
\begin{center}
\begin{tabular}{l r@{.}l r@{.}l r@{.}l r@{.}l r@{.}l r@{.}l r@{.}l}
\hline\hline
 & \multicolumn{2}{c}{Estimate} & \multicolumn{2}{c}{SE} & \multicolumn{2}{c}{$t$-statistic} & \multicolumn{2}{c}{$p$-value} \\
 \hline
$\alpha_0$ & $0$ & $097$ & $0$ & $115$ & $0$ & $861$ & $0$ & $403$\\
$\overline{\Delta}\widehat{Z}_{DMPFC}$& $0$ & $851$ & $0$ & $526$ & $1$ & $619$ & $0$ & $126$\\
$\overline{\Delta}\widehat{Z}_{aINS(r)}$& $-1$ & $506$ & $0$ & $550$ & $-2$ & $737$ & $0$ & $015$\\
$\overline{\Delta}\widehat{Z}_{aINS(l)}$& $-1$ & $126$ & $0$ & $379$ & $-2$ & $967$ & $0$ & $001$\\
\hline\hline
\end{tabular}
\end{center}
\caption{Risk attitude regressed on the average response for all $19$ subjects; $R^2=0.47$, $adjusted$ $R^2=0.36$.}
\label{tab:reg}
\end{table}
 Summary statistics of the model defined in (\ref{eq:risk}) are reported in Table \ref{tab:reg}. Surprisingly, we report that the DMPFC factor, though significantly activated, does not carry explanatory power for risk preferences. This finding, among others, goes far beyond classical fMRI analysis done within the GLM framework and highlights the flexibility and advantages of our approach. Furthermore, the aINS, both left and right regions, are picked up by the model and reported $p$-values are remarkably smaller than $0.05$. Overall, the explanatory power is satisfactory despite the simplicity of linear relation and the noisy nature of the studied panel data (for both, BOLD signal and risk weights). We obtain $R^2=0.47$ and $adjusted$ $R^2=0.36$. The regression fit is depicted in Figure \ref{fig:regs}.
Dropping out of the insignificant terms in (\ref{eq:risk}) yields:
\begin{equation}
\phi_{i}=\alpha_{2} \cdot \overline{\Delta}\widehat{Z}^i_{aINS(l)}+\alpha_{3}\cdot \overline{\Delta}\widehat{Z}^i_{aINS(r)}+\widetilde{\varepsilon}^{i}. \label{eq:risk1}
\end{equation}
The simplified model achieves $R^2=0.37$, $adjusted$ $R^2=0.30$ and the $p$-values are $0.03$ and $0.02$ for $\overline{\Delta}\widehat{Z}_{aINS(r)}$ and $\overline{\Delta}\widehat{Z}_{aINS(l)}$, respectively. Figure \ref{fig:regs} shows the regression fit. In this setup subject risk aversion depends only on the average reaction to the stimulus in the aINS regions. This setup, consisting only of activated (see Table \ref{tab:zscore}) and significant BOLD cluster statistics is kept in the reminder of the analysis.

\subsection{Risk Attitude Forecasting}\label{pred}
\begin{figure}
\begin{center}
\begin{tabular}{cc}
  \includegraphics[width=0.48\textwidth]{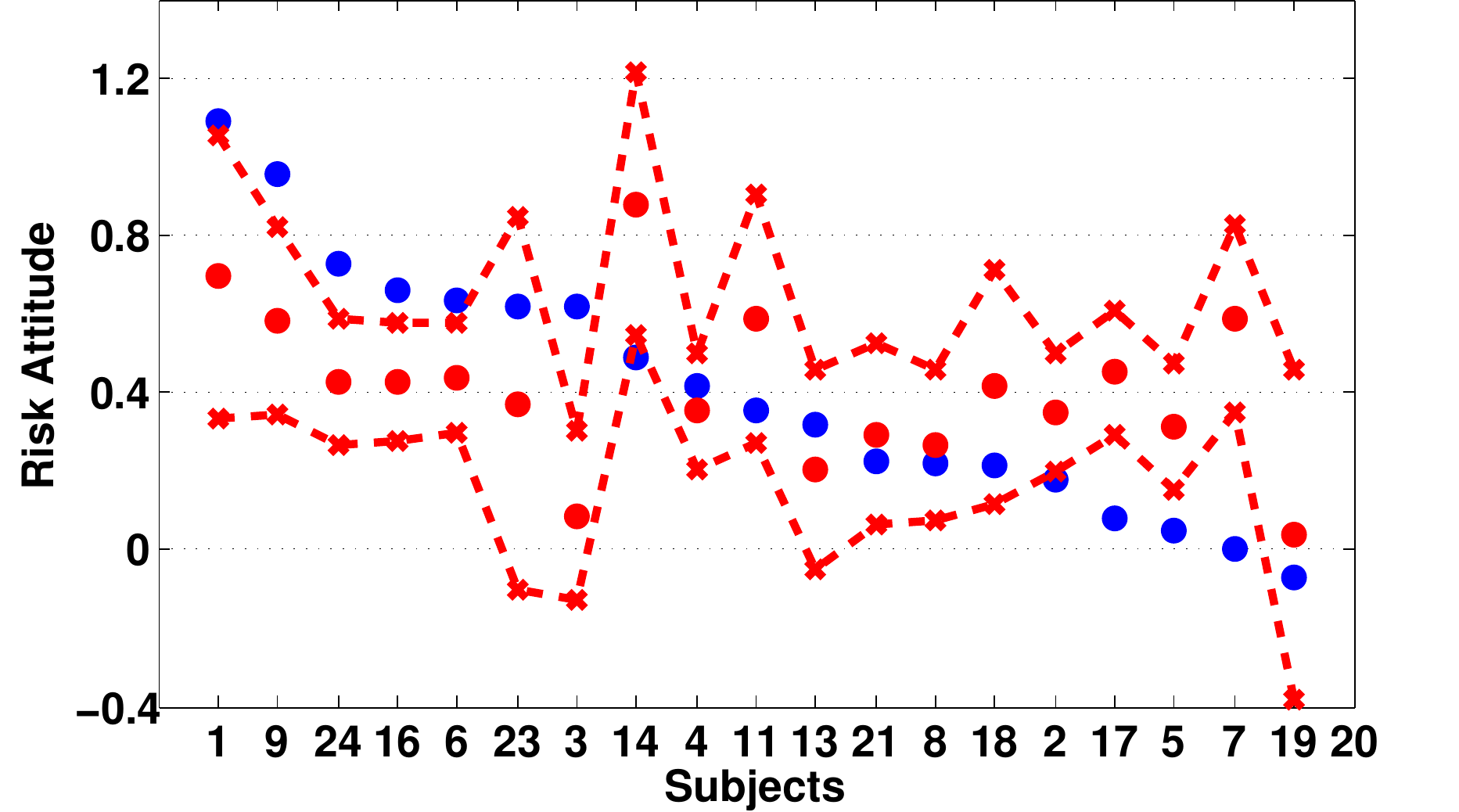}&\includegraphics[width=0.48\textwidth]{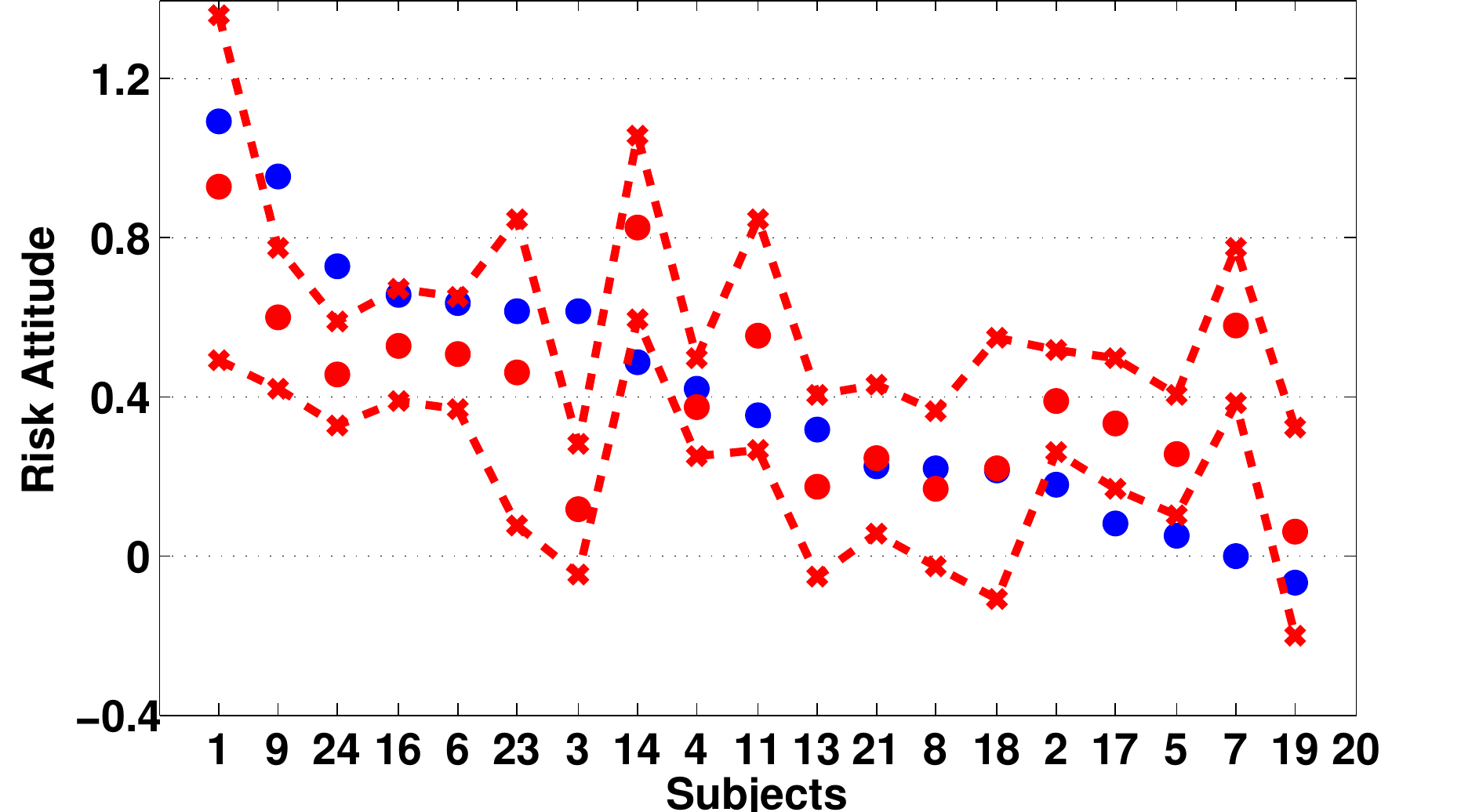}
	\end{tabular}
\end{center}
\vspace{-0.15cm}
\caption{Predicted risk preferences by the model given in (\ref{eq:risk1}) for the average $\overline{\Delta}\widehat{Z}$ and the weighted average $\overline{\Delta}_w\widehat{Z}$: left, right panel, respectively. Information extracted from the aINS BOLD signal; $w=(0.38, 0.41, 0.16, 0.05)^\top$.}
\label{fig:pred}
\end{figure}

The regression results presented in Table \ref{tab:reg} indicate that the DMPFC factor is not significant and does not carry explanatory power for $\phi_{i}$. Thus, the regression setup, stated in (\ref{eq:risk1}) is used to predict the subject risk attitude based only on the information extracted from BOLD signal in aINS. For each subject $i=1,\ldots,19$ its information is excluded from the regression analysis and the model (\ref{eq:risk1}) is re-estimated. Plugging-in the neural low-dimensional representation:  $\overline{\Delta}\widehat{Z}_{aINS(l)}^{i}$ and $\overline{\Delta}\widehat{Z}_{aINS(r)}^{i}$ to the new model predicts the risk weight $\phi_{i}$ and the out-of-sample performance is shown in Figure \ref{fig:pred}. Seven predicted risk attitudes, out of $19$, lie out of $95\%$ prediction confidence intervals and the absolute average forecasting error is $0.257$. One could expect that the proposed statistics $\overline{\Delta}\widehat{Z}$ is not the best univariate projection of the hemodynamic response to the stimulus. To overcome some possible deviations in the HRF peak's location we apply the weighted average reaction to the stimulus denoted by a weighted average reaction: $\Delta_{w}\widehat{Z}^i_{r} = \sum_{\tau=1}^{4}{w}_{\tau}(\widehat{Z}_{r+\tau}^{i}-\widehat{Z}_{r}^{i})$, with $\sum_{\tau=1}^{4}{w}_{\tau}=1$. Thus, observations after stiumuls are weighted with unknown weights $w_{\tau}$. The procedure introduced before is repeated for $\Delta_{w}\widehat{Z}^i=\frac{1}{256}\sum^{256}_{r=1}\Delta_{w}\widehat{Z}^i_{r}$ and the weights are found by minimizing the absolute average forecasting error. The optimal weights $w=(0.38, 0.41, 0.16, 0.05)^\top$ are derived by Monte Carlo simulation with $10000$ iterations and the new absolute average prediction error is $0.202$. The prediction fit is reported in Figure \ref{fig:pred}. In this setup the first $3$ observations (up to $6$ seconds after stimuli) exhibit a remarkably higher impact than the $4$-th one.  

The neural predictions of risk attitudes, though satisfactory, do not perfectly match risk weights derived from subjects' investment decisions. A plausible explanation from a statistical point of view would be the simplicity of linear relation, inhomogeneity of studied subjects and above all, the noisy nature of the data. Nevertheless, we are convinced, that the neural processes underlying investment decisions and corresponding risk preferences are a far more complex phenomenon and go beyond the aINS and DMPFC only. Our statistical methodology is constrained here by the experiment setup that, naturally, cannot capture all brain reactions and allows only to estimate a proxy of "true" risk preferences by risk-return model. Though the activation is reported by the benchmark testing procedure, we suspect additional brain regions to contribute to investment decisions (e.g., \cite{mohr:2010b}) not identified in this fMRI study. This goes beyond the scope of this paper and deserves further research.    

\section{Discussion}\label{discussion}

We have presented a novel method for analyzing fMRI data based on cluster units: CEAD. In the first step the clusters are derived via the NCUT algorithm as contiguous groups of voxels and there are no further constraints concerning the shape and spatial structure. This data-driven approach makes use of the correlation between neighboring voxels and therefore ensures a co-movement of the BOLD signals within cluster. This property of "anatomic" homogeneity pays off when temporal information carried by each cluster has to be extracted. Derived functional connectivity maps are a starting point of analysis. In the estimation-step the DSFM method is applied on each cluster and serves here as a dimension reduction technique. It serves as a filter of the noise and only extracts the common temporal information: the signal (i.e., joint reaction to the stimulus). This semiparametric approach can handle various specifications of noise observed at the voxel level and yields favorable results in comparison to simple averaging over voxels \citep{Heller}. It is a model-free technique that derives complete spatiotemporal information from brain regions. In the activation-step, the extracted signal is further studied for experimental responses. Our local-dynamic representation yields similar results as traditional GLM analyses. The high accuracy of the model plays an important role when possible task-related effects are subtle and local. Our approach ensures a simplicity of neural interpretation and addresses the key limitations of the benchmark method GLM. In the decision step the CEAD method allows for any model-free analysis of spatiotemporal ROI's information. 

We apply the CEAD methodology to study neural systems that underlie decision making under risk. In particular, investment decision is a complex process of valuation and comparison of possible choices with unknown outcomes. Risk attitude is a crucial metric that influences the subjective value of investment. In this paper we analyzed an fMRI experiment with $19$ subjects. Each subject was scanned during multiple ID tasks and a series of $1400$ images of $91\times 109 \times 91$ voxels are investigated here. Using our methodology we decomposed individual brains into sets of $1000$ spatially disjoint factors and factor loadings $\widehat{Z}_{t}^{i,c}$, $i=1,\ldots,19$ and $c=1,\ldots,1000$. Derived spatiotemporal representation is subject-specific and possible variations in functional brain structure are addressed. Therefore we ensure high accuracy and interpretability of the results. Extracted $\widehat{Z}_{t}$ are tested for activation in the GLM (mixed-effects model) framework. For the studied population we detect significant activation at aINS and DMPFC regions as correlates for risk, already reported in \cite{mohr:2010b}. Our approach yields similar results to the benchmark and is complimentary.

To deepen our understanding of changes in neural activity underlying risk preferences we conducted a model-free analysis. The focus is on those ROIs that show ID-related effects: aINS (left and right) and DMPFC (see Table \ref{tab:zscore}) which have previously been associated with decision making. More precisely, we explore the relation between average reaction to the stimulus in subject-specific loadings $\widehat{Z}_{t}$ representing selected regions. Following \cite{Brown} we construct simple, model-free statistics that capture the peak of HRF: $\overline{\Delta}\widehat{Z}_{aINS(l)}$, $\overline{\Delta}\widehat{Z}_{aINS(r)}$, $\overline{\Delta}\widehat{Z}_{DMPFC}$ and explore their explanatory power on the risk attitude $\phi_i$. The resulting regression model with brain dynamics as regressors achieves $R^2=0.47$. Changes in brain activity represented by $\overline{\Delta}\widehat{Z}_{DMPFC}$ did not carry informative power for risk attitude. Simultaneously, both aINS regions are picked up to be statistically significant and reported $p$-values are $\approx0.01$. We conclude that DMPFC, though activated by the risk of the investment, is not significantly correlated to risk attitudes. Dropping off all irrelevant terms and reestimating the regression model (\ref{eq:risk1}) yields $R^2=0.37$. This parsimonious and informative setup is used to predict the risk attitudes based only on fMRI information. The analysis is further refined adjusting for possible variation of hemodynamic response by adding the weights to the sequence of observations after stimulus.

 We report, that neural predictions of risk attitudes, though satisfactory, do not mimic perfectly risk weights derived from subject investment decisions. One may claim that the applied mean-variance model does not reflect true risk attitudes adequately well and additional measures for subjective expected returns and perceived risk than mean and standard deviation should be introduced. Secondly, the risk preferences and neural responses identified in this study may not cover all the effects and brain reactions. Risk attitude is far more complex and may not be only localized in aINS. Therefore we plan to apply our methodology to a wide spectrum of similar studies for further investigations.

\clearpage
\section{Appendix}
\subsection{Simulation Study}

\clearpage
\makeatletter
\setlength{\@fptop}{0pt}
\makeatother
\subsection{Clustering and Sensitivity Analysis}

%
%
\clearpage
\subsection{Factor Loadings}
%

\begin{table}[h!]
\begin{center}
\begin{tabular}{r | r@{.}l r@{.}l r@{.}l | r@{.}l  r@{.}l r@{.}l}
\hline\hline
 & \multicolumn{2}{c}{aINS(l)} & \multicolumn{2}{c}{aINS(r)} & \multicolumn{2}{c}{DMPFC} & \multicolumn{2}{c}{aINS(l)} & \multicolumn{2}{c}{aINS(r)} & \multicolumn{2}{c}{DMPFC} \\
 \hline
KPSS & $0$ & $035$ & $0$ & $063$ & $0$ & $038$ & $0$ & $044$ & $0$ & $051$ & $0$ & $044$\\
ADF & $-0$ & $128$ & $-0$ & $137$ & $-0$ & $110$ & $-0$ & $185$ & $-0$ & $207$ & $-0$ & $159$ \\
\hline\hline
\end{tabular}
\end{center}
\caption{KPSS, ADF test statistics for estimated  factor loadings aINS(left), aINS(right) and DMPFC $\widehat{Z}_t$; subject $1$ (left panel), subject $19$ (right panel) (KPSS: $H_0$: weak stationarity, critical values at $0.10$, $0.05$, $0.01$ are $0.119$, $0.146$ and $0.216$; ADF: $H_0$: unit root, critical values at $0.01$, $0.05$, $0.10$ are $-1.61$, $-1.94$ and $-2.58$).} 
\label{tab:stationarity}
\end{table}
\clearpage
\makeatletter
\setlength{\@fptop}{0pt}
\makeatother
\begin{table}[!t]
\begin{center}
\begin{tabular}{r|c|c|c}

  \hline   \hline

  & DSFM & Average & GLM\\
	\hline
  & $(-34,18,-8)$ & $(-36,18,-8)$ & $(-32,22,-12)$ \\
	aINS(l) & $4.13$ & $4.08$ & $4.58$ \\
	  & $3\times 10^{-4}$ & $4\times 10^{-4}$ & $3\times 10^{-3}$\\
		\hline
		  & $(34,24,-4)$ & $(36,18,-6)$ & $(40,22,-16)$ \\
	aINS(r) & $4.39$ & $4.21$ & $5.24$ \\
	  & $6\times 10^{-6}$ & $6\times 10^{-7}$ & $3\times 10^{-7}$\\
		\hline
				  & $(6,24,42)$ &  $(4,24,42)$  & $(4,24,24)$ \\
	DMPFC & $4.43$ & $3.88$ & $4.56$ \\
	  & $2\times 10^{-9}$ & $1\times 10^{-8}$ & $3\times 10^{-7}$\\
  \hline   \hline

\end{tabular}
	
\caption{The position of the cluster local maximum, denoted in the MNI (Montreal Neurological Institute) standard at $2$mm resolution, corresponding $Z$-score (middle) and $p$-value (bottom) of activated "risk" clusters during the ID stimuli. Average stands for a mean value over voxels in each cluster (results of the NCUT parcellation with $C=1000$). Analysis done in the FSL (FEAT$/$FLAME) software.}
\label{tab:zscore}
\end{center}
 \end{table}

\cleardoublepage

\phantomsection

\addcontentsline{toc}{chapter}{References}

\bibliography{bibfmri}

\end{document}